\documentclass[
 amsmath,amssymb,
preprint,%
]{revtex4-1}

\usepackage{graphicx}
\usepackage{dcolumn}
\usepackage{bm}

\usepackage[utf8]{inputenc}
\usepackage[T1]{fontenc}
\usepackage{mathptmx}
\usepackage{etoolbox}

\makeatletter
\def\@email#1#2{%
 \endgroup
 \patchcmd{\titleblock@produce}
  {\frontmatter@RRAPformat}
  {\frontmatter@RRAPformat{\produce@RRAP{*#1\href{mailto:#2}{#2}}}\frontmatter@RRAPformat}
  {}{}
}%
\makeatother
\begin{document}


\title{The effect of Van der Waals interaction on the microstructure of EPD deposits: a simulation study}

\author{Rémi Martin}
\author{Sandrine Duluard}
\author{Céline Merlet}
 \email{celine.merlet@utoulouse.fr}
 \affiliation{Université de Toulouse, Toulouse INP, CNRS, CIRIMAT, Toulouse, France}


\begin{abstract}
Electrophoretic deposition is a method of choice for generating coatings thanks to its ease of implementation and its ability to produce coatings of relatively large thicknesses in a single step process. While this process also benefits from a large number of tunable parameters to adapt the coating to each application (applied electric field, particle concentration, viscosity of the suspension, etc...), such a freedom can lead to the selection of parameters being an overwhelming task. A better fundamental understanding of the microscopic phenomena and mechanisms at play during deposition can provide clues for the more efficient design of optimized coatings. Particle-based models, allowing for the simulation of deposit microstructures for various process parameters, are particularly interesting to get insights in such systems. Nevertheless, such studies are rare and usually do not involve the possibility of self-cohesion between particles, while it seems crucial for the final structure of the deposit. Here, we use particle-based simulations to study the influence of aggregation on the deposit formed for different applied electric fields. We show that the self-cohesion indeed leads to different microstructures, both in the close vicinity of the substrate and in the bulk of the deposit, and relate this to the mechanical signature of the deposits. Our results reveal that at high electric field, the influence of self-cohesion on resulting microstructures essentially vanishes beyond a critical field strength. This marks the transition between a deposition regime affected by aggregation to a regime largely dominated by volume exclusion effects.
\end{abstract}

\maketitle

\section{Introduction}
    Electrophoretic deposition (EPD) is a cost-effective and versatile technique used to deposit a wide range of materials onto a substrate under various tunable conditions\cite{Boccaccini_2002, VanderBiest_1999}.
    In this process, a colloidal suspension is subjected to an external voltage, particles are driven by the resulting electric field and aggregate into a cohesive network on the surface upon local densification.
    The vast space of processing conditions inherent to EPD has enabled its adaptation to a wide range of applications \cite{Besra_2007, Amrollahi_2016}.
    However, optimizing a parameter set to reliably yield the targeted deposit characteristics remains largely empirical and application-specific\cite{Corni_2008}.
    The green deposits, before any subsequent treatment, are composed of loosely packed particles, and the packing fraction of a deposit as well as its porosity, and more generally its microstructure, depend on the deposition conditions.
    This aspect of the deposits still remains to this day particularly challenging to control, as non-trivial interactions emerge between process parameters, as recent efforts in this direction illustrate\cite{Prioux_2020}.
    
    \begin{figure}[htbp]
        \centering
        \includegraphics[width=0.78\linewidth]{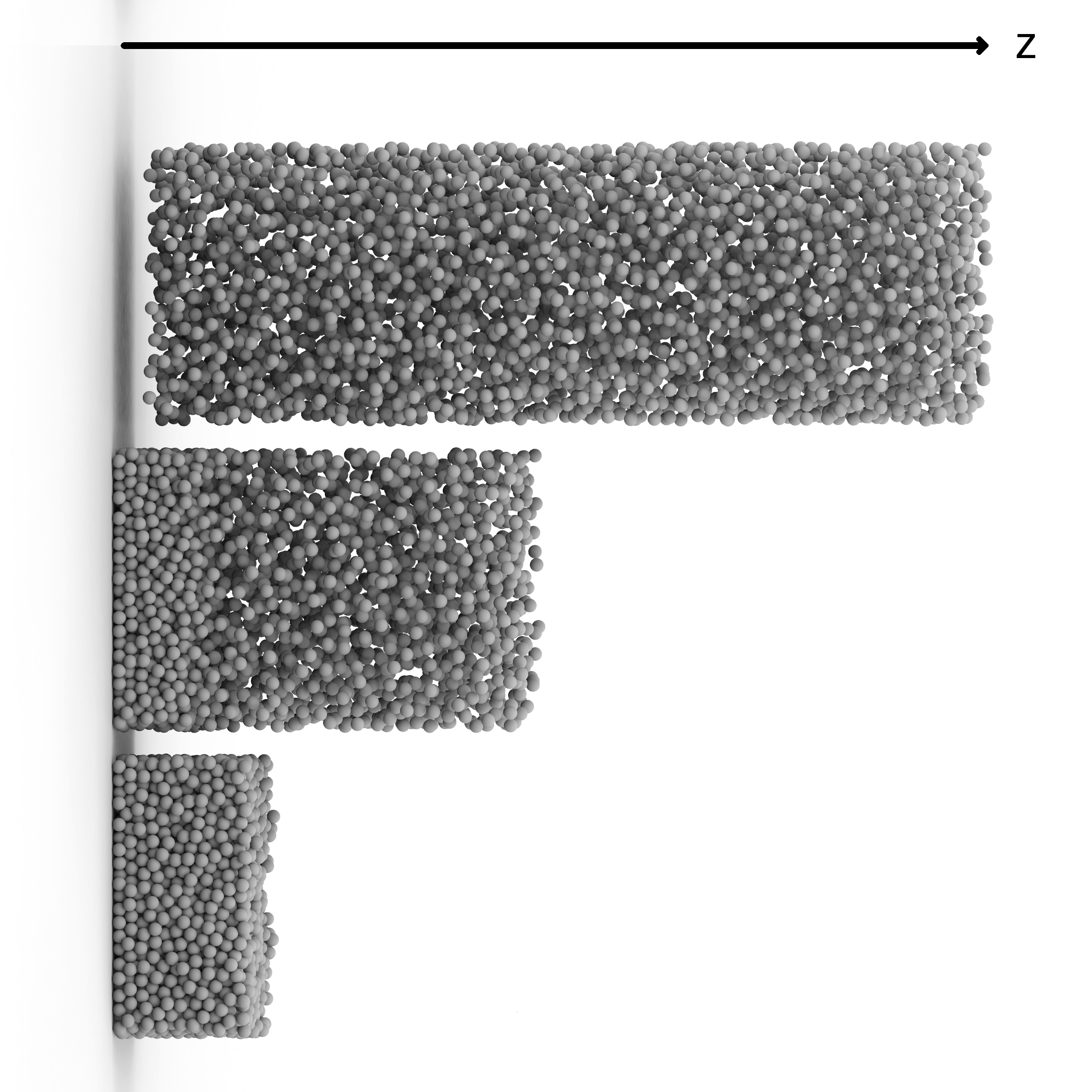}
        \caption{\label{fig: snap_over_time}Snapshots extracted from a particle-based simulation of EPD. Starting from an homogeneous suspension, the particles migrate, over time they accumulate on the substrate and eventually form a deposit.}
    \end{figure}

\begin{figure}[htbp]
        \centering
        \includegraphics[width=0.78\linewidth]{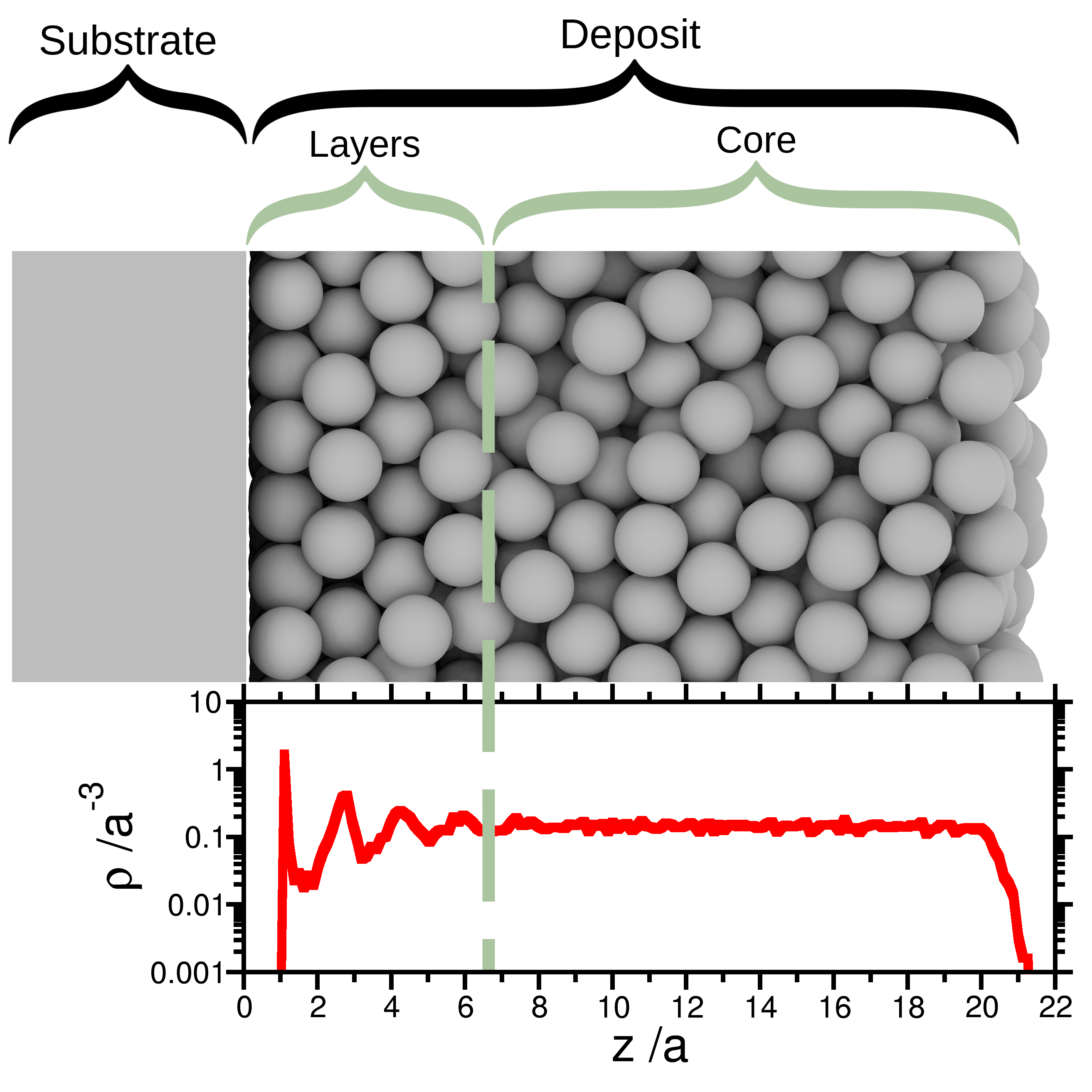}
        \caption{\label{fig: deposit_structure}Structure of the deposit at the end of the simulation showing a layered region, close to the planar substrate on the left, and a more homogeneous core. The structure can be characterized through particle densities (number of particles per volume), given as a function of the distance to the substrate.}
    \end{figure}

    Macroscopic simulations at the continuum level have proven useful for investigating the deposition kinetics and deposit thickness over complex geometries\cite{Verma_2020, SalazardeTroya_2021}.
    Modeling the colloids as a concentration field surrounding the workpiece, and formulating their transport under the influence of an electrical potential enables the prediction of the time evolution of thickness distribution, provided that the film's packing fraction is prescribed.
    By contrast, particle-based approaches capture microstructure characteristics that typically cannot be explicitly described in continuum models, including but not limited to, the deposit packing fraction which naturally emerges from particle simulations.
    Figure~\ref{fig: snap_over_time} illustrates the growth over time of a deposit against a planar surface, Figure~\ref{fig: deposit_structure} shows the typical structure of the resulting deposits.
    However, the current performance of these approaches does not allow for the description of the full range of time and length scales relevant for the deposition process.
    As shown by Giera et al.\cite{giera_mesoscale_2017}, such simulations produce deposits organized at the interface in a layered region, followed by a more homogeneous core.
    Their study highlights the value of this approach for improving our understanding of the relationship between the deposition conditions and the coating microscopic characteristics.
    In what follows, we adopt the methodology developed by Giera et al. by tuning the interaction potential in order to introduce barrier-limited aggregation.
    
    A self-cohesion mechanism is crucial to form a deposit that per\-sists beyond mere particle build-up at the surface.
    In practice, it is often ensured by inter-particle aggregation, linking deposit formation to colloidal stability in the suspension.
    The stability of colloidal suspensions is traditionally understood through the framework of Derjaguin-Landau-Verwey-Overbeek (DLVO) theory~\cite{derjaguin_1941,verwey_1948}.
    Despite the ability of particles to irreversibly aggregate when brought sufficiently close, falling in the deep well of their pair potential, a suspension can be made kinetically stable by the presence of a significant energy barrier\cite{Israelachvili_2011}.
    In their work, Giera et al. used a fixed DLVO pair potential parametrized in such a way to model the deposition of an effective system of non aggregating particles, using an essentially repulsive potential compared to the average thermal energy.
    The DLVO theory, even though its simplicity and accuracy allows for useful modeling and derivations, fails at describing some limiting cases that may appear in some EPD protocols, such as electrically conductive particles, local ion concentration gradients, multi-body effects.
    In an attempt to address this concern, Karnes et al. developed a methodology that aims to overcome some of these limitations, by introducing an adaptive physics, changing on the fly the physical behavior of colloids\cite{Karnes_2024}.
    For instance, their model takes into account particle electric discharge or post-deposition breakups.
    Overall, these previous studies have shown that particle simulations naturally capture some characteristics of the deposit microstructure, but to the best of our knowledge, they have not been used yet to explicitly test the scenarios of barrier-limited aggregation discussed in the EPD literature.
    
    Over the years, several deposition mechanisms have been proposed, involving (i) mechanically induced densification, (ii) electrochemically driven changes in suspension chemistry promoting flocculation, and/or (iii) particle discharge/charge neutralization at the electrode\cite{Sarkar_1996}.
    In the present work we study the pressure-driven mechanism, in the limit where the electric field is intense enough to initiate aggregation without the need for the pressure exerted by upper layers.
    For that purpose we carry out particle simulations of EPD where aggregation is represented by a fixed DLVO pair potential, independent of space and time, and we extract microstructure observables based on the particle and bond distributions.
    We investigate the repercussions of barrier-limited aggregation on the microstructure of EPD deposits in the high-Péclet regime where transport is largely dominated by drift, as the Péclet number corresponds to the ratio between electrophoretic velocity and bulk diffusivity of particles. 

    In a DLVO-type description, the gap between particles bound in the primary minimum lies close to the crossover where short-range van der Waals (vdW) attraction is balanced by even shorter-range steric repulsion.
    Since the steric term varies sharply over sub-nanometric separations, resolving the dynamics becomes a stiff numerical problem.
    In general, particle simulations of colloids that do not account for aggregation omit the vdW term, and it is then very common to treat steric repulsion using a softened repulsive potential, such as the Weeks-Chandler-Andersen (WCA) potential\cite{weeks_role_1971}, which constitutes a standard model system for repulsive colloids\cite{filion_simulation_2011}.
    However, those softened potentials are challenging to make compatible with vdW interaction, because softening the repulsion increases its spatial extension almost immediately far beyond the very short range of the vdW attraction, effectively overwhelming it.
    In our work we decided to use a stiff steric repulsion potential in order to investigate a pair potential featuring a deep well and a significant barrier, at the expense of a very short time step.
    
    Since the systems we model follow Henry's mobility equation\cite{Giera_mesoscale_2015}, we used particularly intense electric fields to keep calculation times reasonable, allowing particles to travel the distance from the extremity of the simulation domain all the way to the substrate.
    The regime we studied is thus largely drift-dominated, the thermal agitation acting as a mere perturbation of the dynamics.
    This regime can be considered as a limiting case rather than a direct point of comparison with experimentally relevant conditions.
    
    If aggregation generally plays a central role in deposition, by influencing the deposition mechanisms it also influences the deposit microstructure.
    This influence is certainly of a different nature in EPD under more experimentally relevant conditions, where diffusion plays a primary role in the deposition dynamics.
    In our work however, studying the effects of aggregation on the deposit formation allows us to highlight some of the mechanisms by which this influence is exerted.
    We study the structure of the deposits formed mainly through quantities derived from the particle space distribution, as well as the aggregate bond space and orientation distribution.
    
    The study is structured as follows.
    In section \ref{methods} we describe the simulation protocol and analyses.
    In section \ref{results} we present and discuss our results first looking at the global effects of aggregation on the deposit structure, then focusing on the core region and finally addressing its specific influence on the interfacial layering.

\section{\label{methods}Methods}

\subsection{Model}

    We model EPD of monodisperse suspensions as assemblies of identical spherical colloid particles, with radius $a$, driven towards a perfectly flat substrate by a homogeneous external electric field directed along the $z$-axis. 
    The simulation box has periodic boundary conditions imposed along the $x$ and $y$ directions, and a wall is present at $z = 0$ representing the substrate.
    The particles are modeled as finite-size rigid bodies with positions and orientations evolving according to Newton's equations.
    In addition to the external electric field, particles experience pairwise colloid-colloid interactions, and dissipative interactions with an implicit solvent.
    
    \subsubsection{Pair potential}
    
        The pair potential models inter-colloid interactions using the classical DLVO framework as the sum of a repulsive long-range screened electrostatic potential $U_e$, a mid-range attractive vdW potential $U_{vdW}$, and a close-range steric repulsive potential $U_r$: 
        \begin{equation}
            U_{DLVO} = U_{r} + U_{vdW} + U_{e}.
        \end{equation}
        The screened electrostatic pair potential is modeled as an exponentially decaying Yukawa potential
        \begin{equation}
            U_{e} = \lambda_DA_Y\cdot\mathrm{e}^{-\frac{h}{\lambda_D}},
        \end{equation}
        where $h$ is the surface-surface distance.
     As detailed in Giera et al.\cite{elimelech_part_dep, giera_mesoscale_2017}, the prefactor $A_Y$ and the Debye length $\lambda_D$ are related to experimentally relevant quantities:
        \begin{equation}
            \lambda_D = \sqrt{\frac{\epsilon k_B T}{2(z_{\pm}e)^2\rho_\pm}}
        \end{equation}
    
        \begin{equation}
            A_Y = \frac{32\pi a\epsilon}{\lambda_D}\left(\frac{k_BT}{z_{\pm}e}\right)^2 \tanh^2\left(\frac{z_{\pm}e\zeta}{4k_BT}\right)
        \end{equation}
        The electrostatic term is thus ultimately parametrized by the temperature $T$, the solvent dielectric permittivity $\epsilon$, the total ion number density far from any colloid $\rho_\pm$ (sum of both anions and cations), the valence $z_{\pm}$ of these ions (the electrolyte is supposed to be symmetrical), and the zeta potential $\zeta$ of the colloid particles. 
        Note that we used a corrected version of the prefactor $A_Y$ by reintroducing a missing constant 4 in Giera's $\tanh$ term\cite{elimelech_part_dep}.
        It leads to a dramatically reduced electrostatic repulsion amplitude and a potential that features a close-range potential well (see Figure~\ref{fig:potentials}).
        
        The vdW and steric potentials are calculated following Everaer's generalization of Hamaker's formulation for the interaction potential of two spheres, each constituted of Lennard-Jones (LJ) particles\cite{Everaers2003}.
        These potentials are parametrized through two quantities, $\sigma_{LJ}$, the radius of the LJ particles, and $A_{cc}$, the Hamaker constant of the spheres.
        The quantity $A_{cc}$ plays a role similar to that of $\epsilon_{LJ}$ for LJ particles, and the two quantities are related through $A_{cc} = \epsilon_{LJ}4\pi^2\rho_{LJ}^2\sigma_{LJ}^6$, where $\rho_{LJ}$ is the number density of LJ particles inside the volume of the spheres:
        \begin{eqnarray}
            U_r &&=  \frac{A\,\sigma_{LJ}^6\,16a^6 \nonumber}{4725\, (h+2a)^8 (h^2 + 4ah)^7}\\
            &&\times\left[491520a^{10} + 2416640a^9h + 5392384a^8h^2\right.\nonumber\\
                      &&+ 7200256a^7h^3 + 6376832a^6h^4 + 3911712a^5h^5\nonumber\\
                      &&+ 1674456a^4h^6 + 490560a^3h^7 + 93660a^2h^8\nonumber\\
                      &&+ \left. 10500ah^9 + 525h^{10}\right],\\
                      \nonumber \\
            U_{vdW} &&= - \frac{A_{cc}}{6} \left[ \ln\left(1-\frac{4a^2}{(h+2a)^2}\right)\right.\nonumber\\ 
            &&\left. + \frac{4a^2h^2+16a^3h+8a^4}{h^4+8ah^3+20a^2h^2+16a^3h} \right].         
        \end{eqnarray}
        
    \subsubsection{External potential}

        A similar Hamaker potential is used for the interaction between particles and the substrate by replacing the radius of one of the spheres with infinity. The potential then becomes a function of the $z$ coordinate of the particle:
        \begin{equation}
            u_{r}^{wall} = \frac{A_{cc}\,\sigma_{LJ}^6 \cdot }{7560}
            \left( \frac{7a-z}{(z-a)^7} + \frac{7a+z}{(z+a)^7} \right),
        \end{equation}
        \begin{equation}
            U_{vdW}^{wall} = - \frac{A_{cc}}{6}\cdot 
            \left[\frac{2za}
            {z^2-a^2}+ 
           \ln\left(\frac{z-a}{z+a}\right)\right].           
        \end{equation}
        The particles experience an additional force constant $\mathbf{F}_E$, as the electric field $\mathbf{E}$ is constant in time and space:
        \begin{equation}
            \mathbf{F}_E = q_{eff}\mathbf{E}
        \end{equation}
        The effective charge $q_{eff}$ of the particles is adjusted so that the electrostatic force matches the hydrodynamic resistance an isolated particle would experience in an electrophoretic steady-state motion, as described by Giera\cite{Giera_mesoscale_2015}:
        \begin{eqnarray}
            q_{eff} &=& \frac{a\pi\epsilon\zeta}{24\lambda_D^6}
            \left(a^5\lambda_D - a^4\lambda_D^2 - 10a^3\lambda_D^3 + 6a^2\lambda_D^4 \vphantom{\int}\right.\\
            &&\left.+ 96\lambda_D^6 + a^4\left(12\lambda_D^2-a^2\right)\mathrm{e}^{a/\lambda_D}
            \int_1^{\infty}dg \frac{\mathrm{e}^{-ag/\lambda_D}}{g}\right).\nonumber
        \end{eqnarray}

    \subsubsection{Dissipative interactions}\label{sec: dissipative}
    
        The interaction with the implicit solvent is modeled through the Fast Lubrication Dynamics (FLD) algorithm\cite{kumar_thesis_2010}, an approximation of Stokesian dynamics\cite{brady_stokesian_1988}. 
        This method has been developed for modelling pairwise short-range lubrication and long-range hydrodynamic interaction, as well as the associated noise, efficiently and consistently with the fluctuation-dissipation theorem, even at high densities.
        The hydrodynamics of the system are accounted for by describing the particle dynamics using a $6N$-dimensional Langevin equation over both translational and rotational degrees of freedom:
        \begin{equation}
            M\frac{d\mathbf{U}}{dt} = \mathbf{F}^H + \mathbf{F}^B + \mathbf{F}^C,
        \end{equation}
        where $\mathbf{F}^C$ is the conservative force-torque vector, $\mathbf{F}^H$ and $\mathbf{F}^B$ represent respectively the hydrodynamic and random Brownian terms of the interaction between the particles and the surrounding fluid, $\mathbf{U}$ is the joined linear and angular velocity vector, and $M$ is the matrix of inertia for both translational and rotational degrees of freedom.
        The hydrodynamic forces $\mathbf{F}^H$, together with the stresslets $\mathbf{S}$, can be expressed as:
        \begin{equation}
            \begin{pmatrix}
                \mathbf{F}^H \\
                \mathbf{S}
            \end{pmatrix}
            =
            \mathbf{R}
            \cdot
            \begin{pmatrix}
                \mathbf{U}^{\infty} - \mathbf{U}\\
                \mathbf{E}^{\infty}
            \end{pmatrix}.
        \end{equation}
        This equation relates the forces and torques ($\mathbf{F}^H$), as well as the stresslets $\mathbf{S}$ experienced by the spheres, to the local fluid linear and angular velocity tensor at the centers of the spheres ($\mathbf{U}$), the bulk fluid linear and angular velocities ($\mathbf{U}^{\infty}$) and the bulk fluid rate of strain tensor $\mathbf{E}^{\infty}$, through a configuration-dependent resistance tensor $\mathbf{R}$.
        In his implementation,\cite{kumar_thesis_2010} Kumar splits the resistance tensor into a sum of two terms,
        \begin{equation}
            \mathbf{R} = \mathbf{R}_0 + \mathbf{R}_{\delta},
        \end{equation}
        a formulation that is close in spirit to Stokesian dynamics.
        The tensor $\mathbf{R}_{\delta}$ corresponds to the near-field contribution to the hydrodynamic resistance. 
        It arises from lubrication terms, asymptotic approximate solutions of the hydrodynamic resistance, obtained for a pair of particles separated by an arbitrarily small gap.
        It can be calculated pairwise, it is frame invariant, and each of its components can be found in the work of Kumar et al.\cite{kumar_thesis_2010}
        
        The essence of the FLD algorithm essentially consists in expressing the other term $\mathbf{R}_0$, representing the far-field hydrodynamic interaction, as a diagonal isotropic tensor that takes the following form:
        \begin{equation}
            \mathbf{R}_0 = 
            \begin{pmatrix}
                6\pi\eta a f^0_{FU}(\phi)\mathbf{I} &0 &0 \\
                0& 8\pi\eta a^3 f^0_{T\Omega}(\phi)\mathbf{I} &0 \\
                0 & 0 & \frac{20}{3}\pi \eta a^2 f^0_{SE}(\phi)\mathbf{I}
            \end{pmatrix}.
        \end{equation}
        It depends on the surrounding fluid's viscosity $\eta$, the radius $a$ of the particles, and the functions $f^0(\phi)$ that depend solely on the volume fraction $\phi$. 
        These functions have been empirically fitted by Kumar to match diffusivities obtained from Stokesian dynamics simulations.
        At each time step, all the forces and torques are calculated, and Newton's equations are integrated by a velocity-Verlet scheme, which corresponds to the inertial formulation of FLD (see Kumar's original publication and thesis for the non-inertial formulation\cite{kumar_article_2010, kumar_thesis_2010}).

    \subsection{Simulation details}

\begin{figure}[htbp]
            \centering
            \includegraphics[width=0.75\linewidth]{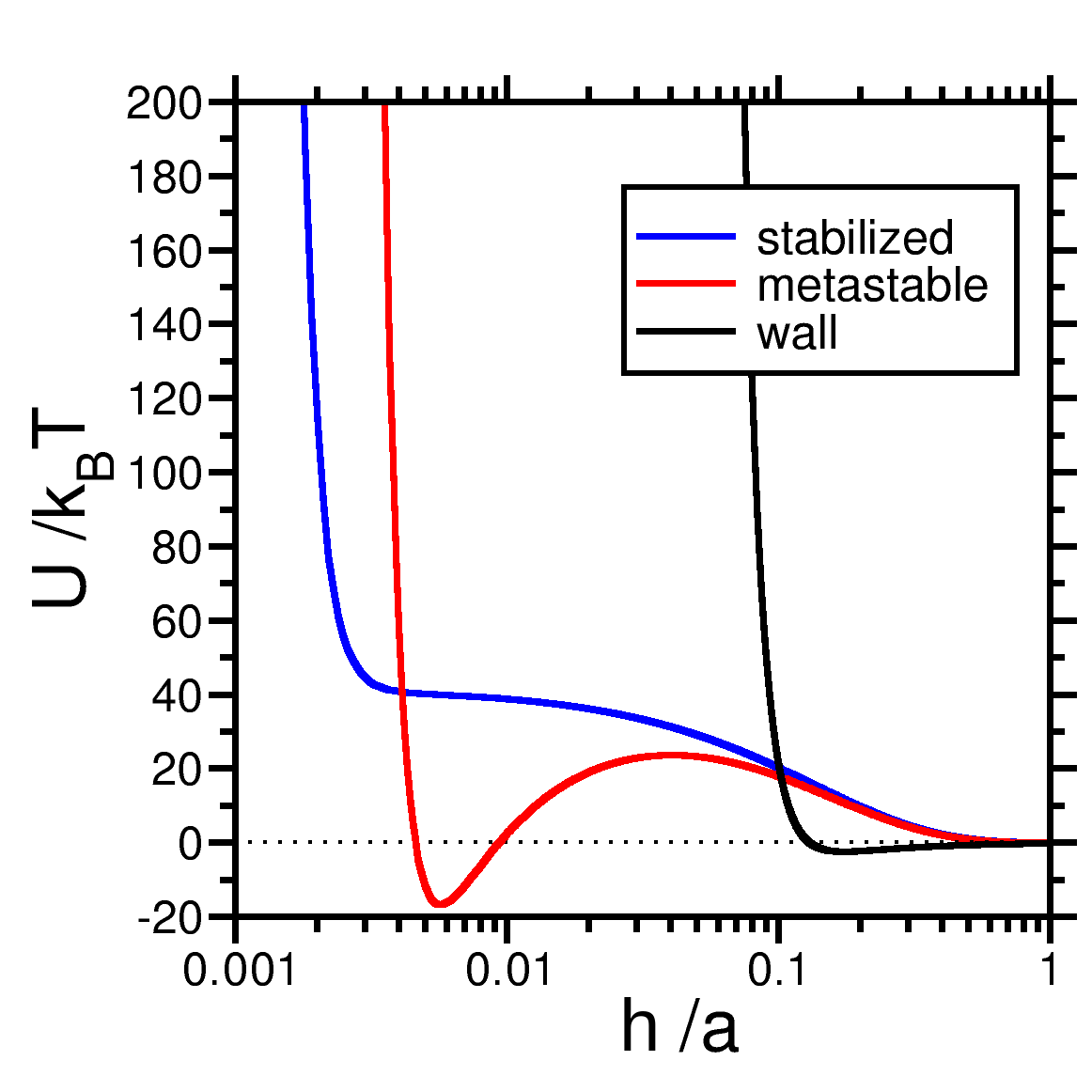}
            \caption{\label{fig:potentials}Pair potentials and wall-colloid interaction potential, as a function of the surface-surface separation~$h$.}
        \end{figure}

        The aim of this work is to compare the structural properties of deposits obtained under varying electric field intensities with two pair potentials: one used as a reference, corresponding to a typical DLVO potential allowing for aggregation, and the other, a fictitious counterpart preventing aggregation (see Figure~\ref{fig:potentials}).
        This fictitious pair potential can be considered as a case in which coagulation is dramatically slowed down in the suspension by the significant potential energy barrier that separates the aggregated state from the free state. It has been constructed from the previous one by simply rendering the vdW term negligible, which removes the potential well altogether, rendering the system stable in the free state, regardless of any exterior conditions. 
        The reference pair potential is called \textit{metastable} while the other is designated as \textit{stabilized}. In the following, we use these terms as convenient labels for the deposits formed from such suspensions, without implying any statement on the thermodynamic stability of the deposits themselves.
         
        The \textit{metastable} and \textit{stabilized} pair potentials converge at a distance that is only a fraction of the particle's radius. 
        For this reason, at sufficiently low particle concentrations, the equilibrium structure of a \textit{stabilized} suspension will arguably match the average structure of a \textit{metastable} uncoagulated suspension. 
        The \textit{stabilized} potential has thus been used to obtain 3 independent equilibrated suspensions with volume fraction $\phi = 0.1$ that serve as starting points for all subsequent EPD simulations.
        To generate initial positions, 5000 particles of radius $a=100~nm$ have been randomly placed in a box of length $L_x=L_y=40a$ along the $x$ and $y$ directions and $L_z=130.9a$ along the $z$ direction. The system has been simulated with periodic boundary conditions along all directions, without any external electric field or any substrate, for a total of $5~10^6$ time steps ($\delta t=20$ps), with the \textit{stabilized} potential, until radial distribution functions (rdfs) have converged.
        
        For EPD simulation, the electric field is introduced as a constant external force field, the periodic boundary condition along $z$ is removed, and a substrate is added at the coordinate $z=0$.
        The 3 equilibrated configurations served as independent starting points for simulating 3 replicas of the EPD process for both \textit{metastable} and \textit{stabilized} systems and for each electric field. 
        For that purpose, the starting configuration is stripped of the particles with coordinate $z<5a$ or $z>L_z-5a$ that might otherwise overlap with the substrate or quickly diffuse out of the box through the higher box side that is not periodic anymore.
        
        A set of 13 different electric fields have been investigated, 12 of which span the 1~MV~m$^{-1}$ to 12~MV~m$^{-1}$ interval, and an additional smaller value of 0.5~MV~m$^{-1}$.
        Consequently, the electrophoretic force ranges from $8.9~10^{-12}$~N to $2.1~10^{-10}$~N, exploring the region of electric fields that are strong enough to induce direct aggregation at contact, as these values remain well above the maximum repulsive force of the DLVO pair potential, $f^{max}_{rep} = 4.9~10^{-12}$~N.
       The Peclet number, expressed as
        \begin{equation}
            \mathrm{Pe} = \frac{|v|}{D}a = \frac{|qE|}{k_\mathrm{B}T}a,
        \end{equation}
        ranges from $\mathrm{Pe}=215$ to $\mathrm{Pe}=5160$, a domain where thermal noise only acts as a small perturbation on the intense drift imposed by the electric field.
        
        The final deposit is not an equilibrium, as such, in order to have comparable deposits, we defined a characteristic time scale in which the simulation time is the same across all the electric field values.
        A single spherical particle in steady-state, interacting solely with the liquid surrounding medium and an external electric field, moves at a constant terminal velocity.  
        The time it takes for the particle to drift by one radius is a natural characteristic time scale for EPD that we can express as: 
        \begin{equation}
            \delta t^{drift} = \frac{6\pi \eta a ^2}{q_{eff}E}.
        \end{equation}
        We performed simulations for a duration $\Delta t = 189\delta t^{drift}$, time at which even the particles that are the furthest away from the substrate had more than enough time to migrate and deposit. 
        The deposit has thus reached a state that is almost steady for all electric fields which allows for comparison between the different microstructures.

        Since we used the inertial formulation of FLD, the fastest dynamic process in the system is the rotational hydrodynamic relaxation of the colloid particles. 
        For an isolated particle in a steady fluid, the characteristic time can be expressed as:
        \begin{equation}
            \tau^h_{rot} = \frac{m}{20\pi \eta a}.
        \end{equation}
        With the parameters used here, it yields $\tau_{rot}^h = 667$ps, which is more than 33 times the time step we used, $\delta t=20$ps.

        \begin{table}
            \caption{\label{tab:table1} Simulation parameters used across all simulations.}
            \begin{ruledtabular}
            \begin{tabular}{lcr}
                Parameter&Value\\
                \hline
                $\delta t$ & $20$~ps\\
                $L_x=L_y$& $40a$\\
                $L_z$& $130.9a$\\
                $a$& $100$~nm\\
                $A_{cc}$& $2~10^{-20}$~J\\
                $\rho_{\pm}$& $5~10^{-4}$~mol~L$^{-1}$\\
                $\epsilon$& $80\epsilon_0$\\
                $T$& 300K\\
                $\zeta$& $20$~mV\\
                $\sigma^{pair}_{LJ}$& $0.01a$\footnote{By contrast, Giera~\emph{et al.} used $\sigma_{LJ}=0.3a$, leading to a steric repulsion extending far enough to cover most of the vdW short range interaction, and thus modelling a non-aggregating system.}\\
                $\sigma^{wall\ }_{LJ}$& $0.3a$\footnote{We used a larger parameter $\sigma_{LJ}$ in the interaction potential with the wall. This explains that the interaction with the wall starts at a greater distance than for a pair of particles.}\\
                $z_{\pm}$& $1$\\
                $d$& 1~g~cm$^{-3}$\\
                $\eta$& 1 Pa~s \\
            \end{tabular}
            \end{ruledtabular}
        \end{table}

    \subsection{Analyses}
        
        The assembly of $N$ particles that constitutes a deposit is characterized by the set $<i,j>$ of pairs of DLVO-bonded particles. 
        Throughout this work, a pair $(i,j)$ is considered bonded if its surface-to-surface separation satisfies $h<0.015a$. 
        For the \textit{metastable} system this threshold corresponds to a decrease of $\sim10~k_{\mathrm{B}}T$ in potential energy relative to the top of the barrier, thus characterizing an irreversibly coagulated state.
        For the \textit{stabilized} system the same threshold is used for direct comparison, although in that case it is only a geometrical criterion and does not correspond to any form of bound state.
        Each pair is accounted for only once in the set, but the choice of $(i,j)$ over $(j,i)$ is arbitrary since bonds are non-polar. 
        We define the lineic distribution of bonds over positions and orientations:
        \begin{equation}
            f(\mathbf{r}, \mathbf{u}) = 
            \sum_{<i,j>} 
            \delta(\mathbf{u}-\mathbf{u}_{ij})
            \int_0^{||\mathbf{r}_{ij}||} d\xi\  
            \delta(\mathbf{r}-(\mathbf{r}_i+ \xi\mathbf{u}_{ij}))
        \end{equation}
        with $\mathbf{r}_{ij} = \mathbf{r}_j-\mathbf{r}_i$ and $\mathbf{u}_{ij}=\frac{\mathbf{r}_{ij}}{||\mathbf{r}_{ij}||}$.
        This distribution sums up to the cumulated bond length contained in the integration volume.
        The choice of the orientation of $\mathbf{u}_{ij}$ is arbitrary since bonds are non-polar.
        All observables considered in this work depend solely on $|\mathbf{u}\cdot \mathbf{v}|$ or  $\mathbf{uu}$, so they are strictly invariant upon the transformation $\mathbf{u}_{ij}\mapsto -\mathbf{u}_{ij}$.
        This arbitrary orientation, therefore, has no effect on the results.
        
        We focus on $f^{\nu}$, which defines the planar distributions over planes perpendicular to the unit vector $\mathbf{e}_{\nu}$:
        \begin{equation}
            f^{\nu}(\mathbf{r}, \mathbf{u}) = |\mathbf{u}\cdot\mathbf{e}_{\nu}| f(\mathbf{r}, \mathbf{u}).
        \end{equation}
        The restriction of this function in a given plane $\Pi_{\nu}$ at altitude $\nu=\mathbf{r}\cdot\mathbf{e}_{\nu}$ is precisely the distribution of bonds that cross this plane, \textit{i.e.} connecting particles from either side of that plane.
        This allows us to naturally expand the $\nu$-profiles of surface-averaged first moments of orientation:
        \begin{subequations}\label{eq:system}
            \begin{align}
                n(\nu) &= \frac{1}{A_{\nu}}
                \iint_{\Pi_{\nu}} d^2A \iint d^2\mathbf{u}\, f^{\nu}(\mathbf{r}, \mathbf{u}),
                \\
                j_{\alpha}(\nu) &= \frac{1}{A_{\nu}}
                \iint_{\Pi_{\nu}} d^2A \iint d^2\mathbf{u}\, f^{\nu}(\mathbf{r}, \mathbf{u})
                \left|\mathbf{u}\cdot\mathbf{e}_{\alpha}\right|,
                \\
                Q_{\alpha\beta}(\nu) &= \frac{1}{A_{\nu}}
                \iint_{\Pi_{\nu}} d^2A \iint d^2\mathbf{u}\, f^{\nu}(\mathbf{r}, \mathbf{u})
                \nonumber\\
                &\qquad \times
                \left(
                \frac{3}{2}\mathbf{e}_{\alpha}\cdot\mathbf{uu}\cdot\mathbf{e}_{\beta}
                - \frac{1}{2}\delta_{\alpha\beta}
                \right).
            \end{align}
        \end{subequations}
        
        The zeroth moment $n(\nu)$ represents the density of bonds that cross the surface at coordinate $\nu$. 
        The first moment of $\mathbf{u}$ is exactly zero since $f(\mathbf{r},\mathbf{u}) = f(\mathbf{r},-\mathbf{u})$, so the appropriate observable here is $j_{\alpha}$, the first moment of $|u_{\alpha}|$ along the direction $\mathbf{e}_{\alpha}$.
        This quantity measures the mean absolute projection of the orientations of plane-crossing bonds along $\mathbf{e}_{\alpha}$.
        In particular, $j_{\nu}$ corresponds to the absolute flux of bonds that cross the plane.
        The second moment $Q_{\alpha\beta}$ of $\mathbf{u}$, however, is nonzero and captures the orientational anisotropy in the plane.
        
        This continuum description allows us to relate these observables to the lineic position-orientational density $f(\mathbf{r},\mathbf{u})$. 
        However, their discrete definitions, strictly equivalent, are more directly useful to configuration analysis. The system is homogeneous and isotropic in the $xy$ plane, so the natural direction we take to calculate profiles is $\nu=z$:
        \begin{equation}
            n(z) = \frac{1}{L_xL_y} \sum_{<i,j>_z} 1
        \end{equation}
        \begin{equation}
            j_z(z) = \frac{1}{L_xL_y} \sum_{<i,j>_z} |u_z|
        \end{equation}
        In this work, we focus on the flux component along the direction $\nu$ because it can be easily related to the tensile strength about the plane $\Pi_{\nu}$:
        \begin{equation}
            \sigma^{max}_{\nu\nu}(\nu) = |f^{max}_{att}|j_{\nu}(\nu),
        \end{equation}
        where $f^{max}_{att}=-2.3\cdot 10^{-10}$~N is the maximum attractive force of the DLVO potential.
        
        To characterize the anisotropy of the bond structure around the $z$-axis, we define the following order parameter:
        \begin{equation}
            S(z) = \frac{Q_{zz}(z)}{n(z)} = \frac{\sum_{<i,j>_z} \left(\frac{3}{2}u_z^2-\frac{1}{2}\right)}{\sum_{<i,j>_z} 1}
        \end{equation}
        Although $S(z)$ is mathematically related to the classical nematic orientational order parameter, it is not defined in a volumetric domain but over the set of plane-crossing bonds, which is more relevant to this confined deposition mechanism.
        However, its isotropic baseline value is nonzero, $S^\mathrm{iso}=\frac{1}{4}$, reflecting the geometrical bias of crossings toward larger values of $|u_z|$.
        It ranges from $-\frac{1}{2}$ when all bonds are parallel to the plane to $1$ when perpendicular.
       
\section{\label{results}Results and Discussion}
    \subsection{\label{results: I}The structure of the deposits}

\begin{figure}[htbp]
            \centering
            \includegraphics[width=\linewidth]{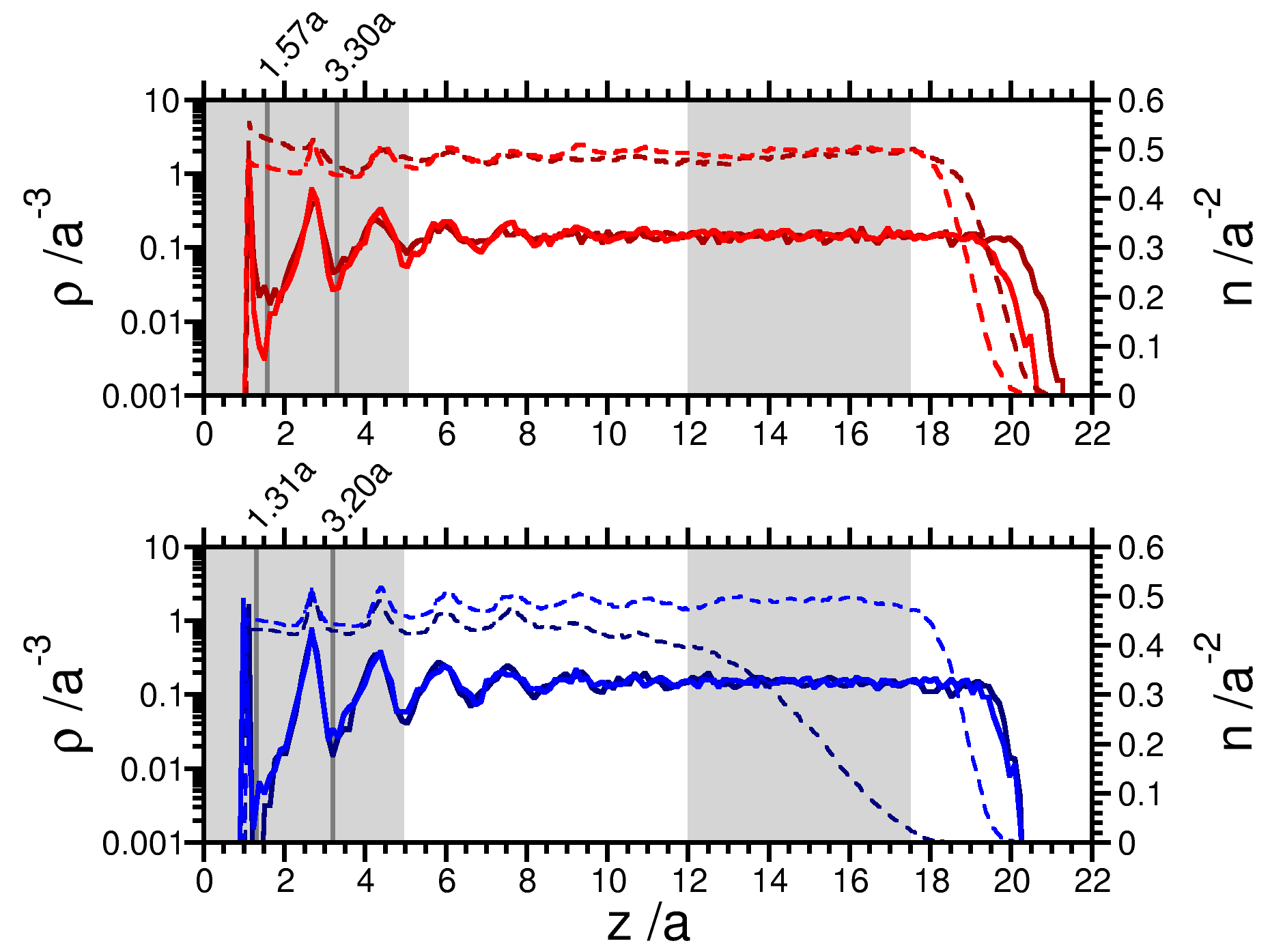}
            \caption{\label{fig: density_profiles} $z$ profiles of particle density $\rho$ (continuous lines) and $xy$ plane-crossing bond density $n$ (dashed lines). 
            \textit{Metastable} deposits are in red, and \textit{stabilized} deposits are in blue. 
            The darker color describes deposits obtained with $E=0.5MV.m^{-1}$, and the lighter ones for $E=12MV.m^{-1}$. 
            The grey region on the right covers the subregion within the core studied in part \ref{results: II}, and the one on the left covers the first 3 layers studied in part \ref{results: III}.}
        \end{figure}

        Both deposits formed from \textit{stabilized} particles (unable to coagulate) and \textit{metastable} particles (subject to barrier-limited coagulation) are typically composed of two distinct regions, as can be seen in the particle density profiles $\rho(z)$ in Fi\-gu\-re~\ref{fig: density_profiles}: a layered interface and a bulk-like core.
        The electric field influences these regions differently: increasing $E$ enhances layering in \textit{metastable} deposits (layering around 10$a$ distance as compared to around 7$a$ at lower electric field), whereas it tends to reduce interfacial order in \textit{stabilized} systems (layering around 10$a$ in both cases but large decrease of plane crossing bond density at higher electric field). 
        Moreover, \textit{metastable} deposits progressively approach \textit{stabilized} microstructures as $E$ increases, which we quantify below.

\begin{figure}[htbp]
            \centering
            \includegraphics[width=0.7\linewidth]{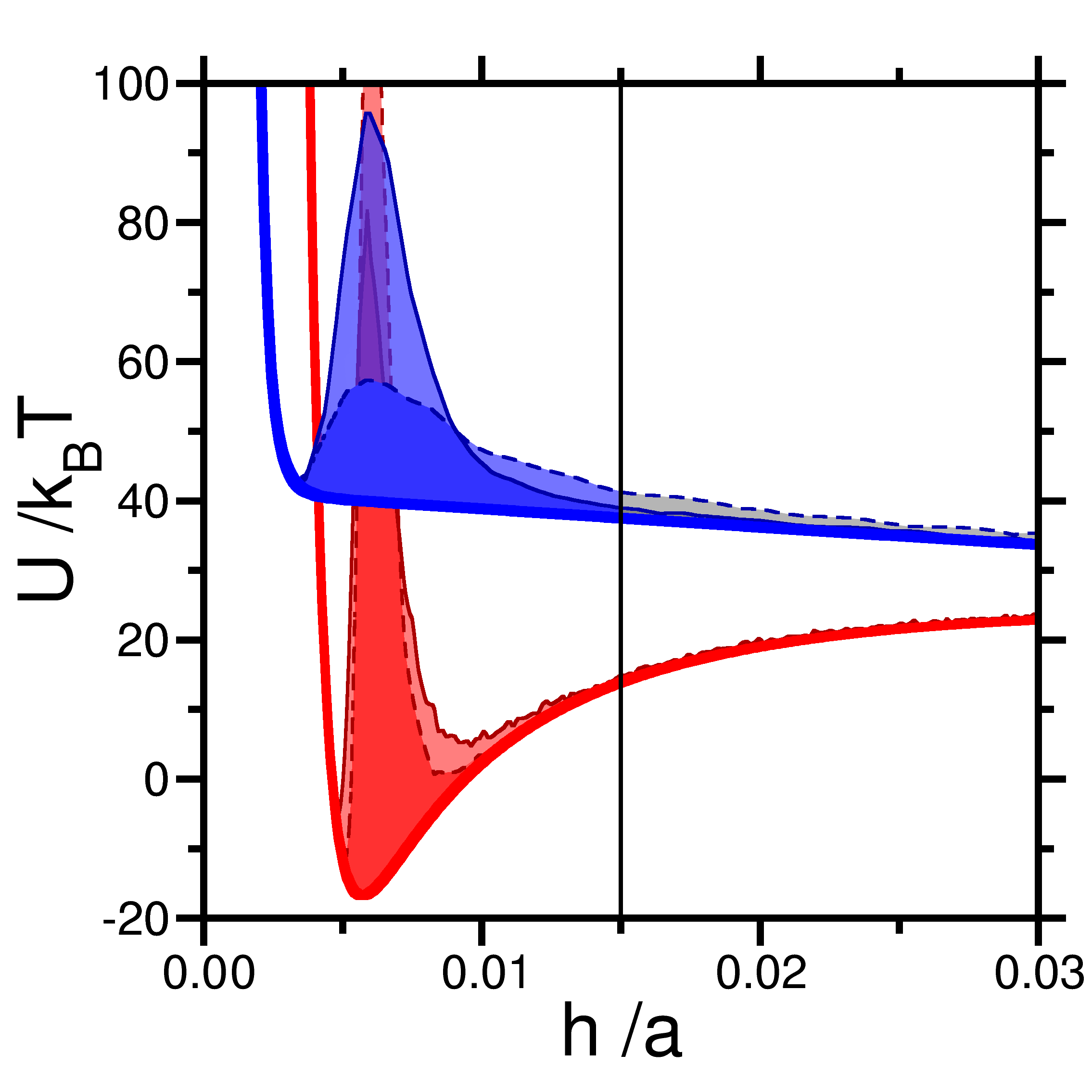}
            \caption{\label{fig: first shell}Pair potentials (thick lines) and first rdf peak, averaged over the bulk (thin lines). The rdf represented are those obtained for $E$~=~0.5~MV~m$^{-1}$ (dashed lines), and $E$~=~12~MV~m$^{-1}$ (continuous lines). \textit{Metastable} deposits are in red, and \textit{stabilized} deposits are in blue. The vertical line at $0.015a$ represents the distance criterion within which a pair is considered as bonded.}
        \end{figure}

        \medskip
        We first examine the core region ($\sim10a$ to $\sim20a$). 
        The particle density $\rho(z)$ appears roughly constant across this domain (solid lines, Figure~\ref{fig: density_profiles}), around 0.1~a$^{-3}$, suggesting quasi-homogeneous packing.
        Yet the density $n(z)$ of bonds crossing horizontal planes (dashed lines), which probes deposit connectivity, reveals some degree of underlying heterogeneity. 
        For \textit{metastable} deposits, $n(z)$ remains nearly constant.
        In contrast, \textit{stabilized} deposits exhibit a dramatic decay of $n(z)$, which extends more deeply into the core as the electric field weakens, consistent with elastic compression under hydrostatic load.
        The nature of this decay, in terms of changes in local neighbor distances, can be examined using the rdf, averaged over the core (Figure~\ref{fig: first shell}).
        For \textit{metastable} deposits, neighboring particles aggregate within the potential well, and decreasing $E$ narrows the first-neighbor peak without shifting its position.
        For \textit{stabilized} deposits, lacking a binding well, the distribution shifts continuously to smaller separations with $E$.
        At low $E$, the broad rdf peak spans both sides of the bond threshold ($h = 0.015a$, vertical line); at high $E$, compression shifts nearly all pairs below this cutoff. 
        This $E$-dependent shift is observed despite averaging over the core ($12~a$ to $17.5~a$), thereby superposing rdf curves corresponding to different depths.
        It is consistent with elastic compression governed by the hydrostatic pressure, which increases with both applied field and depth in the deposit.
        This elastic signature, corresponding to a displacement of first-neighbor positions of only a few percent of particle radius, does not alter the quasi-homogeneous character of the core.
        To quantify any residual departure from homogeneity, a linear regression of the core density profile reveals a weak but significant spatial decrease, of around $\Delta\phi \approx -0.02$, in both \textit{stabilized} and \textit{metastable} deposits.
        Although subtle, the fact that this decrease is observed in both systems suggests slow, irreversible structural relaxation under sustained load, an effect examined in more detail in Sec.~\ref{results: II}.

        \medskip
        For films formed from \textit{metastable} particles, increasing $E$ reduces the deposit thickness from $\sim21a$ to $\sim20a$ and systematically shifts the overall density profile $\rho(z)$ towards that of \textit{stabilized} deposits. 
        This structural convergence is most visible in the layered region at the interface with the substrate (from the substrate to $\sim10a$). At low $E$, \textit{metastable} layers are poorly defined (lower layer density, and more interlayer particles), whereas at high $E$, peak and well heights meet those of \textit{stabilized} deposits under the same field.
        Interestingly, \textit{stabilized} deposits show the opposite trend: decreasing $E$ tends to improve layer differentiation, with slightly sharper density oscillations, higher peaks and lower densities in the inter-layer regions.
     
        Beyond particle densities, plane-crossing bond densities (dashed lines, Figure~\ref{fig: density_profiles}) reveal the mechanical implications of such reorganization.
        In \textit{stabilized} deposits, this profile exhibits the expected pattern. Roughly constant between layers, it corresponds essentially to bonds connecting two consecutive layers; within the layers, the peaks correspond to the in-layer bond network.
        Increasing $E$ uniformly raises $n(z)$, consistent with the elastic compression discussed in the previous paragraph. 
        In contrast, \textit{metastable} deposits at low $E$ show a singular behavior: $n(z)$ decreases with $z$ across the first two interlayer regions, which could be a signature of aggregation-driven structural arrest.
        
\begin{figure}[htbp]
            \centering
            \includegraphics[width=\linewidth]{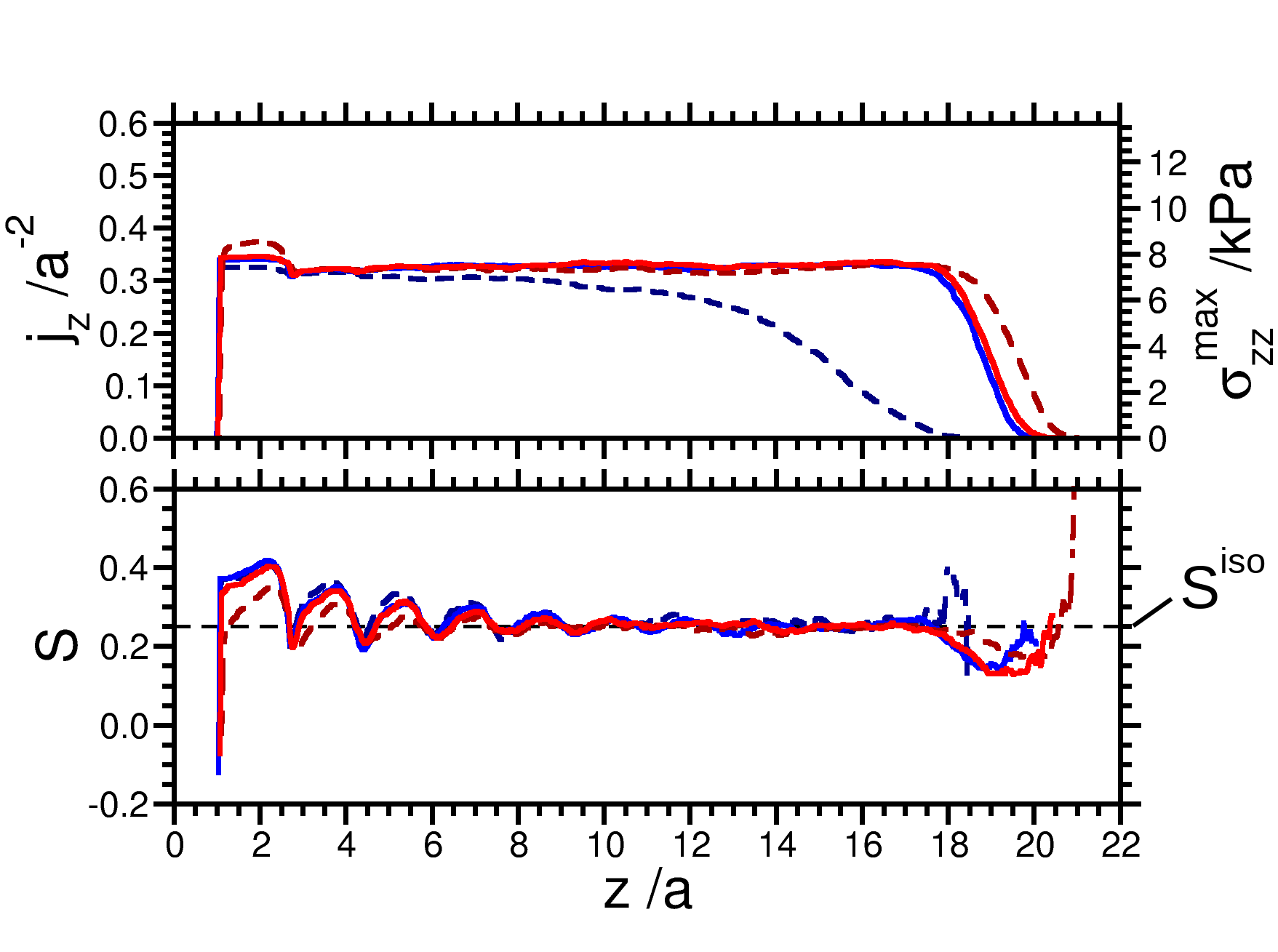}
            \caption{\label{fig: bond profiles}$z$ profiles of bond orientational observables. Flux $j_z$ and the associated maximum failure stress $\sigma_{zz}^{max}$, as well as the orientational order parameter $S$ are represented, for $E$~=~0.5~MV~m$^{-1}$ (dashed darker lines) and $E$~=~12~MV~m$^{-1}$ (continuous lighter lines) for \textit{metastable} and \textit{stabilized} deposits (respectively red and blue).}
        \end{figure}

        Other aspects of the orientation distribution can be assessed from the bond flux $j_z(z)$ and the order parameter $S(z)$ (Figure~\ref{fig: bond profiles}).
        Unlike $n(z)$, which peaks within layers, the bond flux transitions from one interlayer region to another, almost in a stair-like manner, reflecting the vertical connectivity. 
        The vertical connectivity governs resistance to delamination (in-plane bonds, confined to the $xy$ plane, bring negligible contributions to $j_z$).
        Across the core, $j_z$ varies little with $E$ for \textit{metastable} deposits, though in the layered region, the profiles for both \textit{metastable} and \textit{stabilized} systems seem to converge toward the same high-field limit. 
        These contrasting trends in thickness, layering, and bond geometry point to a field-induced transition between two deposition regimes in the \textit{metastable} system, the high-field regime approaching that of \textit{stabilized} particles. 
        This transition is quantified in the following sections.
     
        \medskip
        The bond network geometry described above has direct mechanical implications for the freshly formed deposits. 
        These signatures characterize the intrinsic cohesion of the solvent-filled structure immediately after deposition, prior to any drying or consolidation.
        For \textit{metastable} deposits, aggregation confers intrinsic self-cohesion: these structures form colloidal gels capable of surviving field removal and withstanding subsequent handling.
        In contrast, \textit{stabilized} deposits lack self-cohesion and would redisperse immediately upon field removal. 
        For \textit{metastable} colloidal gels, $j_z$ provides a direct measure of mechanical integrity, as it is proportional to the ultimate tensile stress $\sigma_{zz}^{max}$.
        for both \textit{stabilized} and \textit{metastable} deposits, $j_z$ in the layered region has a step-like shape, which is most clearly seen upon magnification, as shown in the Supplementary Material.
        For \textit{stabilized} deposits, $j_z$ lacks direct mechanical significance but can nonetheless remain useful as a geometric point of comparison.
        For \textit{metastable} deposits however, each step indicates the delamination resistance between consecutive layers: it shows that the layered region near the substrate in the inter-layer $1-2$ exhibits the strongest connectivity, while the coating inter-layer $2-3$ is the weakest, with core values lying intermediate between these interfacial extremes.
        
        Beyond ensuring cohesion, aggregation alters the orientational order of the bond network. 
        As noted above, decreasing $E$ degrades layering in \textit{metastable} deposits while improving it in \textit{stabilized} ones. 
        This different structural response is better quantified by the order parameter $S(z)$: while \textit{stabilized} deposits show little $E$-dependence, \textit{metastable} deposits at low $E$ exhibit $S$ closer to the isotropic reference $S^\mathrm{iso} = 1/4$ throughout the layered interface (both within layers and between them).
        This increase in orientational disorder indicates that, at low $E$, aggregation disrupts the formation of well-organized layers, preventing \textit{metastable} structures from achieving the denser packing observed in \textit{stabilized} deposits.
        
        Finally, we note an intriguing feature at the core-suspension interface. 
        Approximately one diameter before $\rho(z)$ starts decaying, $S(z)$ decreases (preferential in-plane alignment), then reverses closer to the interface (though low bond density there limits statistical significance).
        Such behavior may reflect transient rearrangements following the arrest of deposition caused by the depletion of suspended particles, a hypothesis that motivates future dynamical analysis.
                   
    \subsection{\label{results: II}Average properties in the core}
    
        As established in Sec.~\ref{results: I}, the core exhibits a quasi-homogeneous structure when described in terms of local density profiles.
        Beyond this spatial characterization, the core provides a natural framework to interpret the deposit microstructure in temporal terms, as the outcome of competing non-equilibrium processes occurring upon and after deposition.
        All core-averaged quantities reported in this section are computed within the interval $12a<z<17.5a$, which defines a nearly homogeneous region of the deposit for the different deposition conditions.
        Accordingly, core-averaged properties, such as volume fraction, pore size, and (for \textit{metastable} deposits) bond statistics, can be treated as representative of the entire core region.
        For \textit{stabilized} deposits, however, bond-derived quantities lack mechanical significance.
        The threshold $h < 0.015a$, while useful for geometric comparison in section~\ref{results: I}, does not correspond to true cohesive binding.
    	Consequently, core-averaged bond statistics are reported only for \textit{metastable} deposits in what follows.
         \begin{figure}[htbp]
            \centering
            \includegraphics[width=0.75\linewidth]{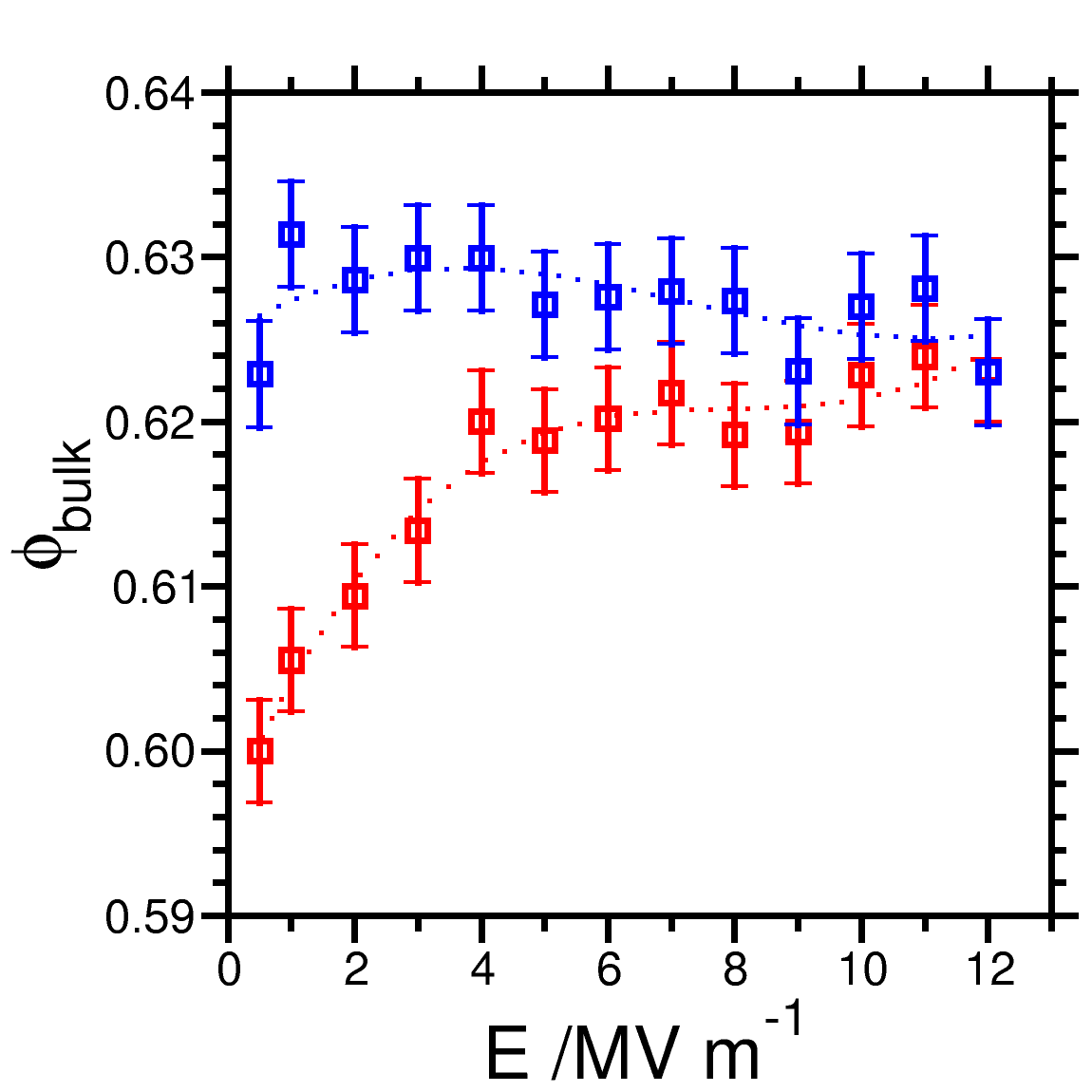}
            \caption{\label{fig: core vol frac}Average volume fraction in the core of the deposit ($12a<z<17.5a$) as a function of electric field. \textit{Metastable} deposits are in red, and \textit{stabilized} deposits are in blue.}
        \end{figure}  
        
        \medskip
        The core represents a quasi-stationary state in which particle mobility is strongly reduced by crowding effects, with an additional effect of inter-particle bonding in \textit{metastable} deposits. This reduced mobility prevents relaxation toward the denser configurations achievable by full equilibration.	
        Core-averaged volume fraction $\phi_\mathrm{core}$ (see Figure~\ref{fig: core vol frac}) reveals contrasting behaviors: for \textit{stabilized} deposits, $\phi_\mathrm{core}$ shows no significant $E$-dependency and remains around $\sim0.628$, well below close packing ($\phi\approx 0.74$) whereas in \textit{metastable} deposits a similar maximum packing value is obtained at high $E$ but decreases to 0.6 at low $E$.
        
        Upon reaching the deposition front, particles experience rapid densification since crowding dramatically reduces particle mobility. At any given depth, this manifests as a sharp transition from dilute suspension to dense deposit, with densification continuing until mobility drops below the threshold required for further structural relaxation.
        This kinetic arrest mechanism resembles the hard-sphere glass transition. While the \textit{stabilized} particles interact via screened electrostatics rather than hard-core repulsion, the absence of attractive wells renders all contact configurations energetically similar—making volume fraction the dominant control parameter. 
        Classical work on these systems shows the existence of a fluid-solid first-order phase transition\cite{alder_1957}, where a low-density fluid coexists with a high-density crystal at constant pressure, in the range between the freezing and the melting points, respectively at $\phi_f = 0.494$ and $\phi_m = 0.545$\cite{phan_1996}. When hard spheres are not given enough time to relax to the thermodynamic equilibrium, a path-dependent out-of-equilibrium glass transition occurs. In the literature on hard-sphere glasses, kinetic arrest typically occurs above $\phi\approx0.58-0.62$, depending on preparation protocol and relaxation times\cite{parisi_2005}. Spherical colloids can display similar phase and glassy behavior upon densification, and can be crystallized or form supercooled liquids and glasses in the same ranges of packing fractions\cite{hunter_2012}.
        
        The directed constraint given by the electric field, the planar confinement imposed by the substrate, and the interfacial nature of the deposition front make the detailed mechanisms of kinetic arrest in our EPD simulations somewhat different from that of bulk suspensions undergoing isotropic strain. However, while the directed nature of EPD differs from isotropic densification protocols, the observed $\phi\approx 0.628$ and loss of mobility are consistent with colloidal glass formation. The deposition mechanism is then interpreted within a glass-transition-like framework, in which kinetic arrest at the deposition front limits further compaction of the core. This path-dependent kinetic arrest implies that core density should vary with $E$, as faster deposition reduces relaxation time at the front. However, statistical uncertainty in $\phi_\mathrm{core}$ obscures this trend.
        
        As shown in section~\ref{results: I}, \textit{metastable} deposits converge toward \textit{stabilized} structures at high $E$. At these electric fields, vdW interactions are increasingly negligible compared to the external electrostatic force, and thus deposition of \textit{metastable} particles lies in the same \textit{glass transition regime} as for deposits from \textit{stabilized} particles, and $\phi_\mathrm{core}$ is similar (0.624).
        At low $E$, however, the volume fraction of the core drops down to $\phi\approx0.6$. 
        Here, aggregation from vdW interactions upon deposition becomes a decisive factor in limiting the volume fraction, in addition to steric hindrance imposed by neighboring particles.
        In what follows the terms \textit{glass transition regime} and \textit{aggregation regime} are used to qualify deposition regimes based on the dominant mechanisms limiting structural compaction.
        \begin{figure}[htbp]
            \centering
            \includegraphics[width=0.75\linewidth]{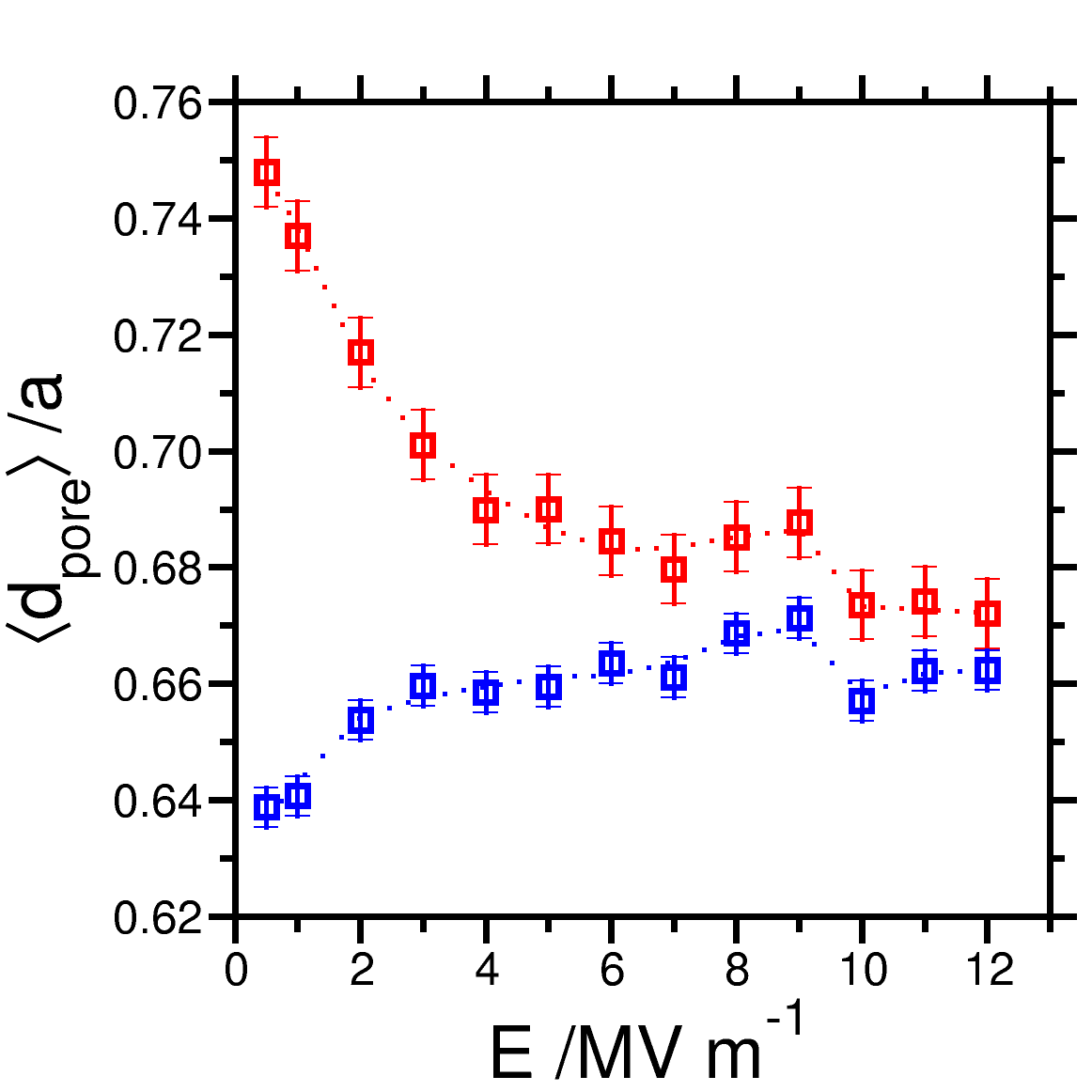}
            \caption{\label{fig: pore diameter}Average pore diameter in the core region of the deposits ($12a<z<17.5a$) as a function of electric field. \textit{Metastable} deposits are in red, and \textit{stabilized} deposits are in blue.}
        \end{figure}

        \medskip    
        To further explore the evolution of the microstructure with applied electric field, a more sensitive probe has been tested with the pore size distribution (Figure~\ref{fig: pore diameter}).
        Whereas the bulk density is unchanged for \textit{stabilized} deposits, the core-averaged pore diameter shows a modest yet significant increase with electric field. This trend reflects the deposition mechanism underlying the glass transition regime: higher $E$ shortens the residence time at the front, reducing the time window for structural relaxation. Consequently, more open structures (larger pores obtained at high $E$) are also structures that are kinetically arrested earlier, before achieving denser configurations.
        At high $E$, \textit{metastable} deposits exhibit pore sizes approaching those of \textit{stabilized} deposits.
        Crucially, this convergence is not limited to the average, but concerns the entire pore size distribution (see Supplementary Material), a robust signature of the glass transition regime. 
        At low $E$, \textit{metastable} deposits exhibit significantly larger pores, consistent with reduced $\phi_\mathrm{core}$. This reflects the influence of aggregation on configurational arrest: bonded networks resist compaction more effectively than purely repulsive interactions. In this aggregation low $E$ regime, bonds formed near the front are strong enough to resist electric field driven structural collapse. 
        Despite their distinct microscopic mechanisms (kinetic arrest in \textit{stabilized} and \textit{metastable} deposits at high $E$ and aggregation-assisted arrest in \textit{metastable} ones at low $E$) both regimes ultimately freeze configurations at the deposition front.
        
        After the phenomena occurring primarily at the deposition front, the deposit is not static: it continues to relax over time, slowly altering the configuration initially imprinted, through a process analogous to glass aging at least in the case of \textit{stabilized} deposits.
        While most core-averaged observables combine contributions from formation and post-deposition evolution, the spatial density gradient observed along the core in both deposits provides a particularly direct signature of slow structural relaxation.
        In a steady deposition regime, the structure imprinted at a given depth mainly reflects the time elapsed since the passage of the front and the subsequent mechanical environment, notably the depth-dependent load.
        A gradual densification with depth therefore naturally emerges from relaxation processes acting over increasing residence times and hydrostatic pressures, consistent with the negative density gradient measured in the core for both \textit{metastable} and \textit{stabilized} deposits ($\Delta\phi \approx -0.02$).
        Remarkably, this spatial variation is comparable in magnitude to the total change in the core-averaged volume fraction observed in \textit{metastable} deposits when varying the electric field. 
        Since the latter was argued above to reflect a change in the deposition regime, this suggests that formation and relaxation-induced effects may contribute at similar levels to the final structure.
        At the same time, we cannot exclude front-related transients associated with a slow evolution of the deposition front itself during growth.
        In that case, the structure emerging from the front may depend weakly on the stage of deposition, independently of post-deposition relaxation in the bulk.
        Disentangling the respective roles of front evolution and post-deposition relaxation, as well as assessing their impact on core-averaged observables beyond the density gradient, requires a time-resolved characterization of the deposition process, most naturally formulated in a reference frame co-moving with the deposition front.
        Such an analysis lies beyond the scope of the present static study but would constitute a natural extension of this work.
        \begin{figure}[htbp]
            \centering
            \includegraphics[width=0.75\linewidth]{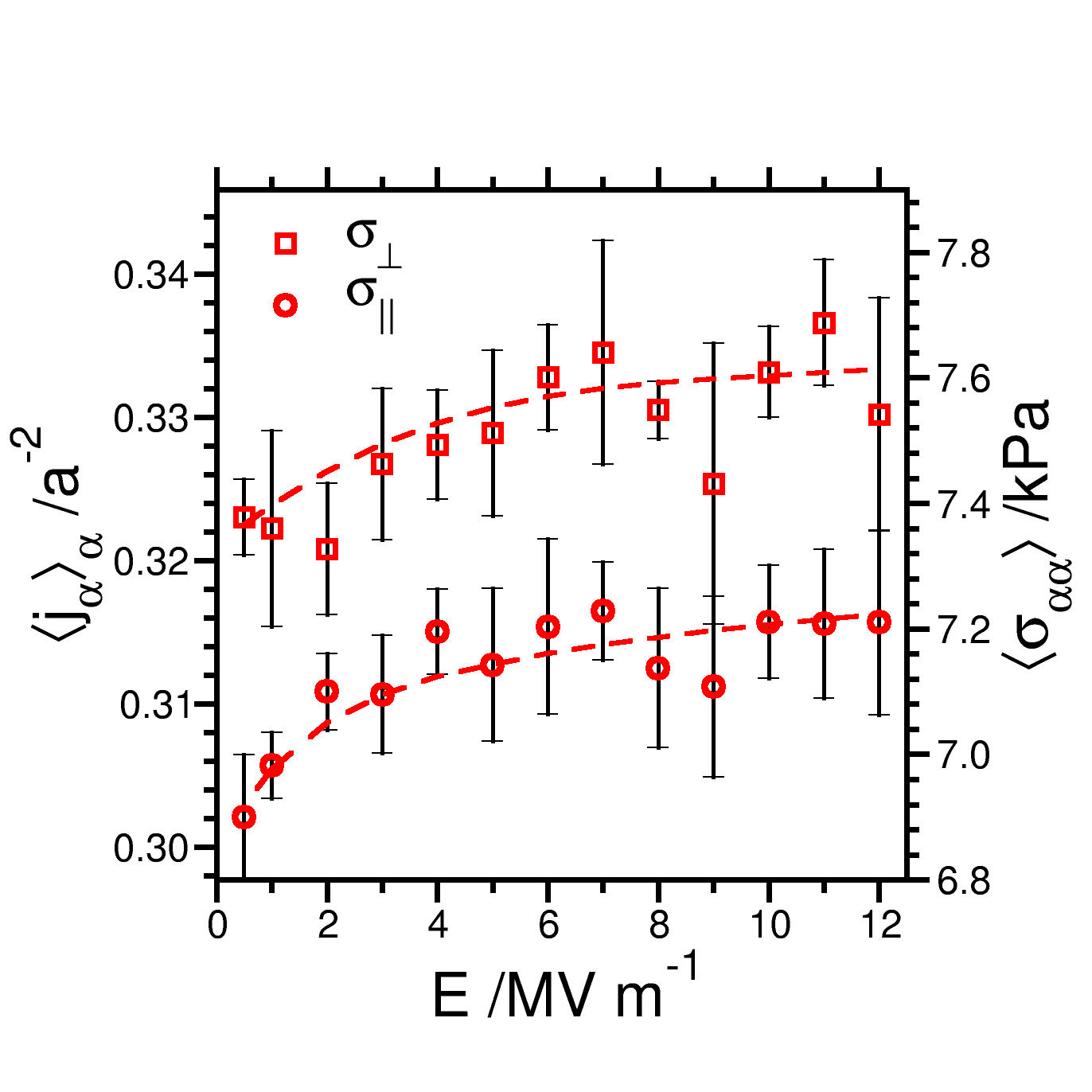}
            \caption{\label{fig: tensile strength}Core-averaged tensile strengths as a function of electric field for \textit{metastable} deposits, along axes perpendicular and parallel to the $xy$ plane (respectively $\sigma_{\perp}$ and $\sigma_{||}$)}
        \end{figure}

        \medskip
        While layered regions govern delamination (section~\ref{results: I}), the core provides a natural setting to quantify the intrinsic cohesive strength of the bonded network, free from layering artifacts. For \textit{metastable} deposits, the mechanical properties can be determined via the calculation of tensile strengths, from the bond flux $j$ along the direction $\nu$.
        
        Core-averaged tensile strengths $\sigma_\perp$ (perpendicular to substrate) and $\sigma_{||}$ (parallel) reveal persistent mechanical anisotropy for all $E$ (see Figure~\ref{fig: tensile strength}), with $\sigma_\perp > \sigma_{||}$.       This anisotropy, although modest in such a quasi-homogeneous region, reflects the directional nature of EPD; bonds formed under electrophoretic driving retain a preferential vertical alignment. 
        In both directions, tensile strength increases with $E$ in the aggregation regime, as stronger fields compress bonded networks more tightly. At high $E$, (glass transition regime), this trend weakens as kinetic arrest progressively supersedes bond formation as the primary limit to densification.
        Together, these mechanical measurements provide a coherent mechanical picture of \textit{metastable} deposits, from the deposit's core down to the substrate interface, discussed in the next section.

    \subsection{\label{results: III}Microstructure of the layered region}
        The layered region at the interface with the substrate is the most structurally and mechanically heterogeneous region of the deposit, likely to be the most fragile region in wet deposits, which is of particular interest in the context of delamination.
        Aggregation noticeably reshapes this region: at low $E$, \textit{metastable} deposits show different layering quality and bond orientation compared to \textit{stabilized} systems.
        We now quantify these structural signatures, focusing on layer density, interlayer bonding, and two-dimensional packing geometry.
        \begin{figure}[htbp]
            \centering
            \includegraphics[width=0.75\linewidth]{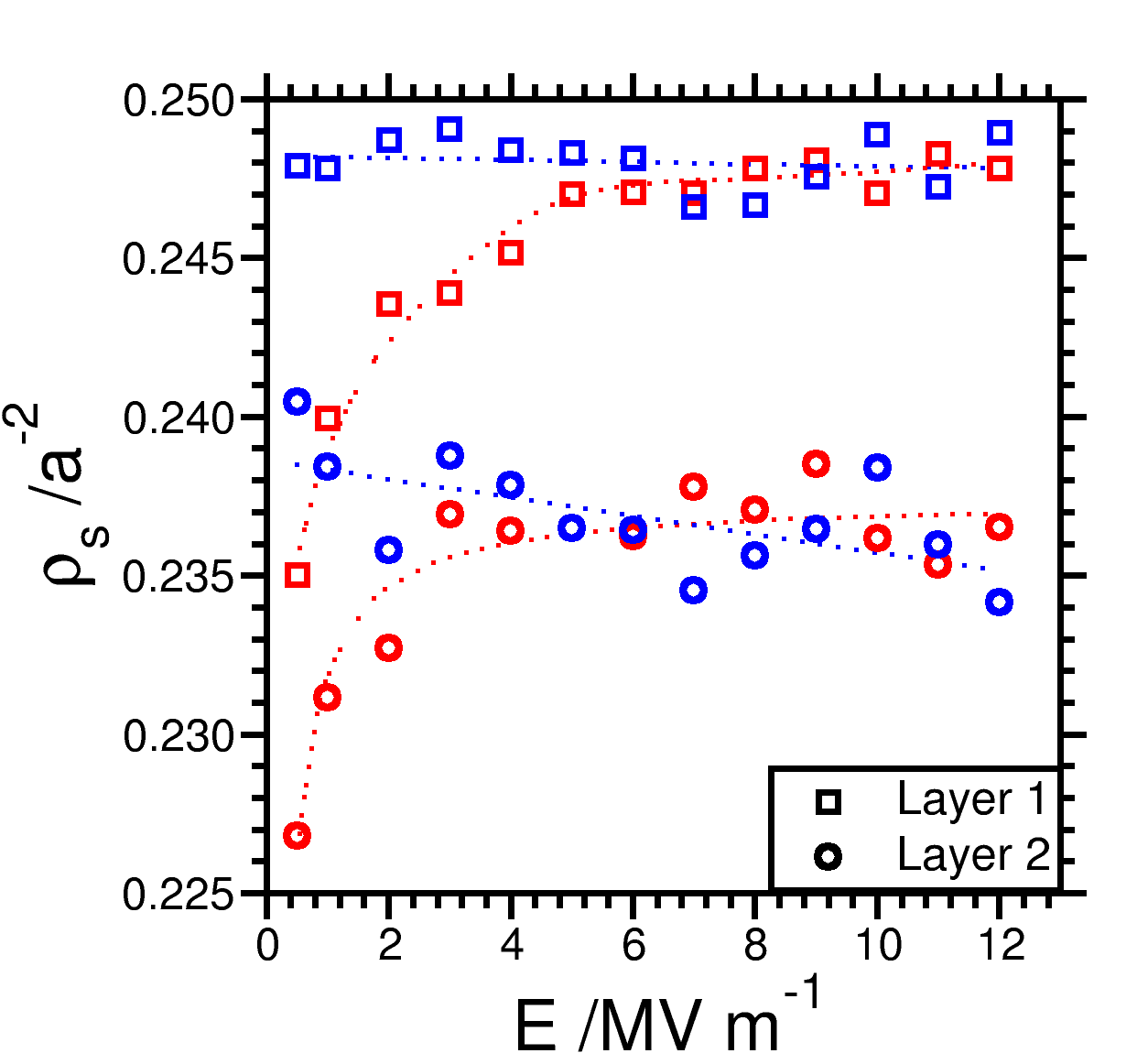}
            \caption{\label{fig: layer densities}Surface density of particles in the first two layers as a function of electric field. \textit{Metastable} deposits are in red, and \textit{stabilized} deposits are in blue. }
        \end{figure}
        
        \medskip
        To assign each interfacial particle to its corresponding layer, we defined the boundaries of the first three layers as the z-coordinates of the two adjacent density minima. While these boundaries show no systematic $E$-dependence within each system, they differ slightly between \textit{metastable} and \textit{stabilized} deposits. We thus use system-specific boundaries, averaged over all values of $E$, which yield average boundary positions of $z/a=1.57,3.30$, and $5.08$ for \textit{metastable} deposits. These layers are slightly thicker than those of \textit{stabilized} deposits, for which the boundaries are set at $z/a=1.31,3.20$, and $4.96$.
        Using these boundary positions, we compute the surface particle densities of the first two layers. As shown in Figure~\ref{fig: layer densities}, the first layer is consistently denser than the second, regardless of the electric field or the aggregative properties of the particles, reflecting the higher disorder imposed by an underlying layer as compared to a perfectly planar substrate.
        For \textit{stabilized} deposits, the layer densities show no significant $E$-dependence whereas   for \textit{metastable} deposits, the transition toward the \textit{glass transition regime} is marked by a progressive merging of the two systems around $E\sim$~6~MV~m$^{-1}$, beyond which the layer densities become indistinguishable from those of \textit{stabilized} deposits.
        In the aggregation regime ($E<6$~MV~m$^{-1}$), however, decreasing the electric field leads to a marked reduction in layer density of these \textit{metastable} deposits. Particle aggregation leads to the formation of bridges, which limit further layer densification, a phenomenon analogous to what has been observed in the quasi-homogeneous environment of the core.
        \begin{figure}[htbp]
            \centering
            \includegraphics[width=\linewidth]{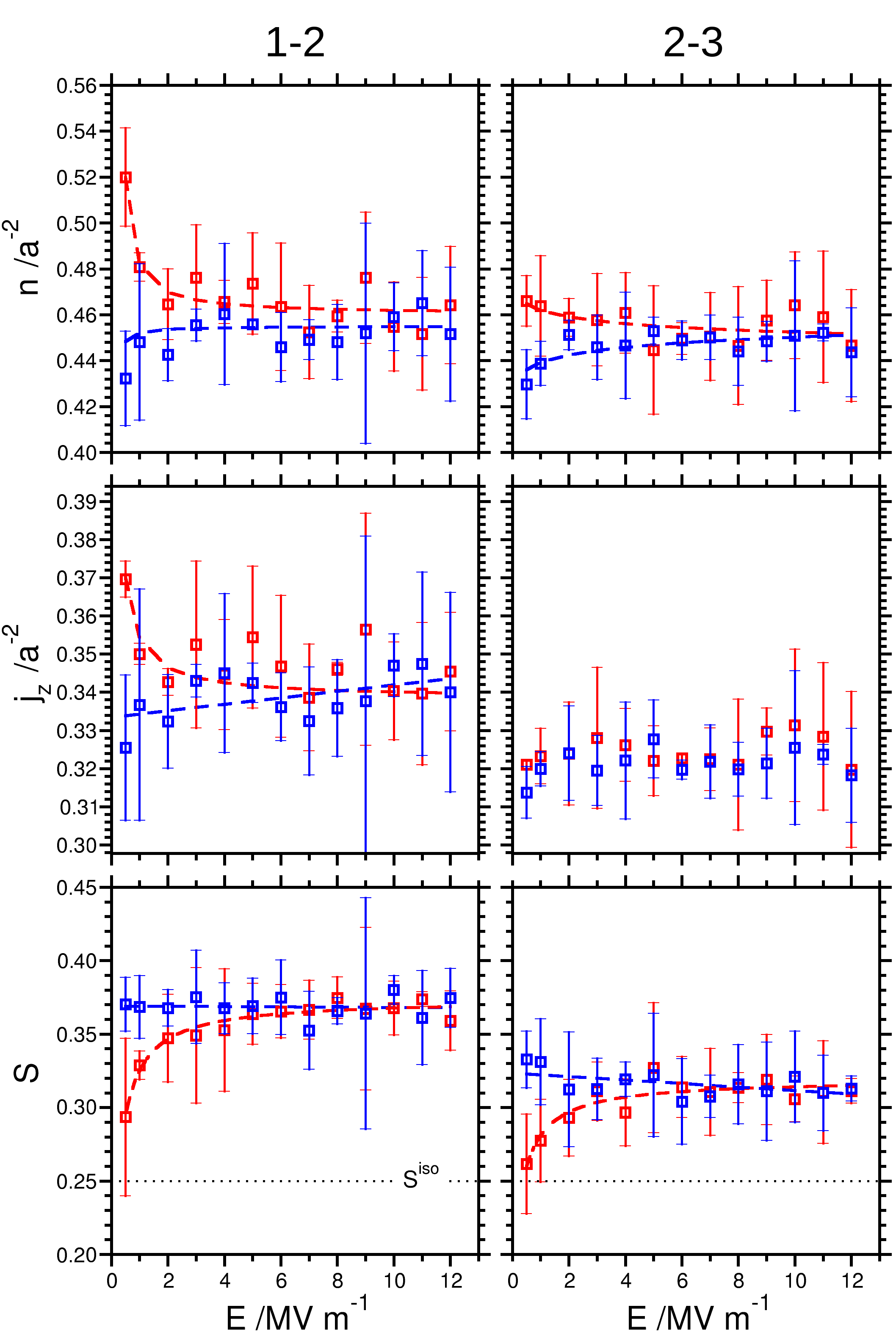}
            \caption{\label{fig: layer bonds}Layer bond surface density $n$, orientational order parameter $S$, and bond flux $j_z$, proportional to tensile strength $\sigma_{zz}=j_z\left|f^{max}_{att}\right|$, with $f^{max}_{att}=-2.3\cdot 10^{-10}$~N, as a function of the applied electric field. Average at the intersection between layers 1 and 2 on the left, and between layers 2 and 3 on the right. \textit{Metastable} deposits are in red, and \textit{stabilized} deposits are in blue. 
            }
        \end{figure}

        \medskip
        At first sight, the reduction in layer density of \textit{metastable} deposits might be expected to result in weaker interfacial cohesion. Bond network analysis, however, reveals a more nuanced picture.
        The plane-crossing bonds quantification approach is particularly well suited for this purpose, as it allows inter-layer bond statistics to be naturally evaluated at the $z$-positions corresponding to the boundaries between adjacent layers.
        The resulting bond density $n$, flux $j_z$, and order parameter $S$ are reported in Figure~\ref{fig: layer bonds} for the $1-2$ and $2-3$ planes.
        \textit{Metastable} deposits exhibit a clear increase in inter-layer bond density $n$ as $E$ is reduced, most prominently in the $1-2$ plane. Quantitatively, a $10\%$ increase in interlayer bond density is associated with a decrease of about $4\%$ in the particle densities of the corresponding layers.  This indicates  a modification in the local particle environment, with a larger number of bonds per particle connecting successive layers.
        The associated bond flux $j_z$, which is proportional to the tensile strength $\sigma_{zz}$, displays a weak increase in the $1-2$ plane at the lowest values of $E$, although the limited statistical significance does not allow for a firm conclusion.
        
        The clear mismatch between the variations of bond density $n$ and bond flux $j_z$ points to a change in bond orientation distribution, as reflected by the order parameter $S$. For all conditions considered, $S$ exceeds the isotropic reference value $S^\mathrm{iso}$, reflecting the preferential alignment of inter-layer bonds along the deposition direction.
        As the electric field is reduced and the bond density increases, however, $S$ decreases toward the isotropic limit. 
        This indicates that the bond orientation distribution tilts towards the $xy$-plane, which compensates for the bond density increase, but possibly not enough to prevent the increase we observe in the bond flux $j_z$, at least for the $1-2$ plane.
        At higher $E$, as the deposition regime of \textit{metastable} particles transitions from \textit{aggregation} to \textit{glass transition} mechanism, the inter-layer bond statistics progressively converge to the one of \textit{stabilized} system (that is \textit{glass transition} driven).
        Taken together, these observations indicate that aggregation promotes a local disorganization of the interfacial layers: particle surface densities decrease, while bond orientations become more isotropic.
        The behavior of bond flux suggests that this aggregation-induced disorder affects inter-layer connectivity in a way that may enhance the effective tensile response of the interface, although this effect remains difficult to quantify conclusively.
         \begin{figure}[htbp]
            \centering
            \includegraphics[width=\linewidth]{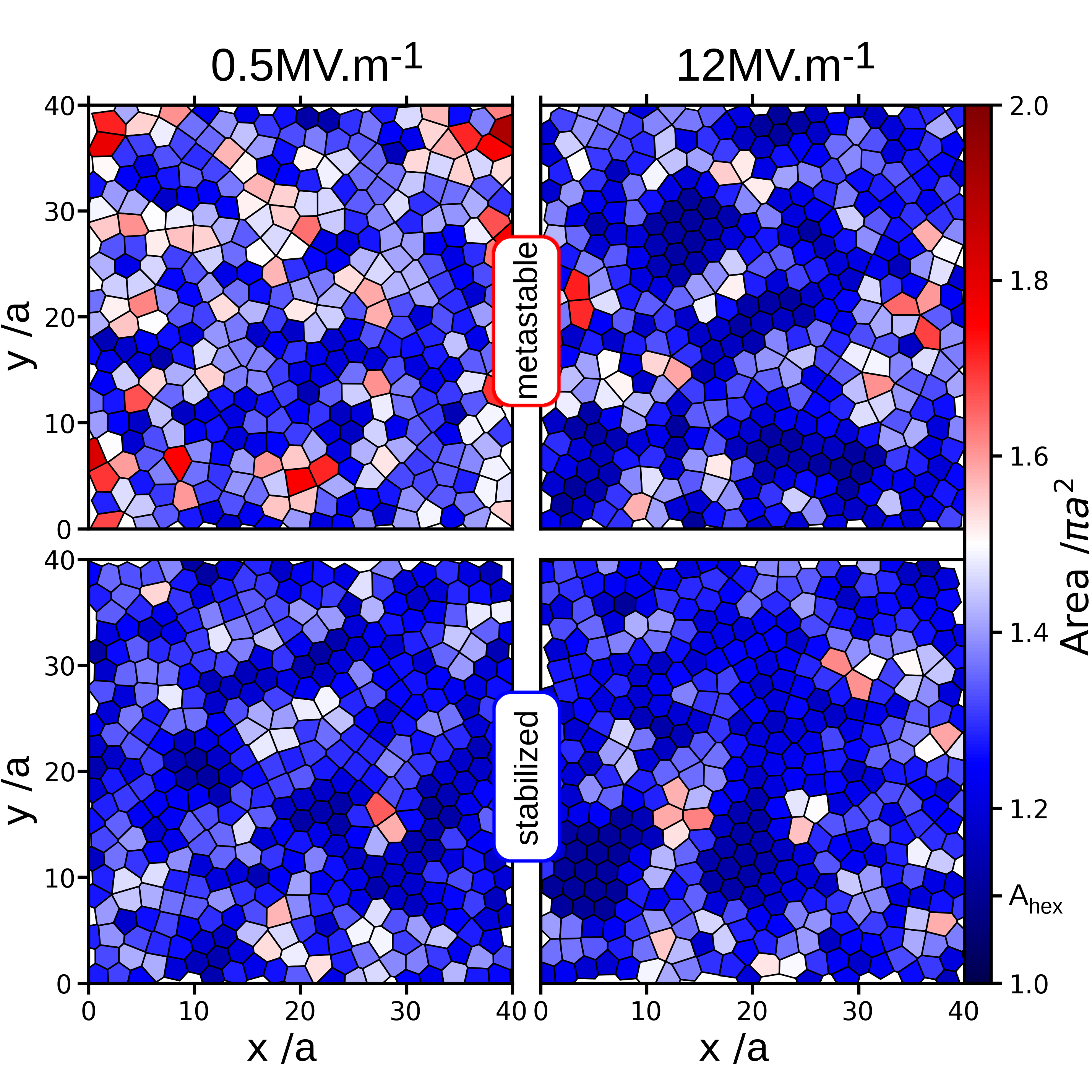}
            \caption{\label{fig: voronoi diag}Voronoi tilings of first layers projected in the $xy$ plane.}
        \end{figure}
        
        \begin{figure}[htbp]
            \centering
            \includegraphics[width=0.75\linewidth]{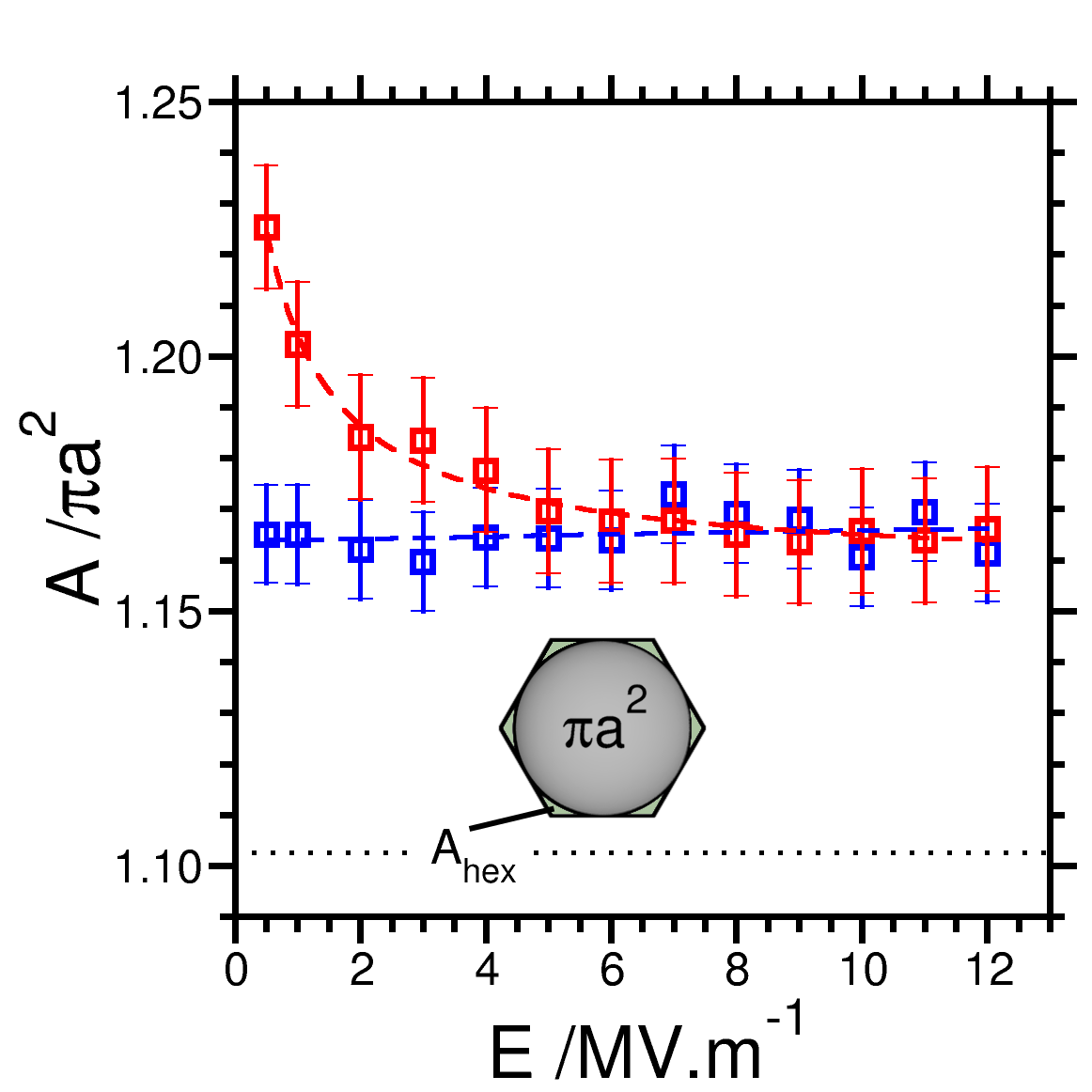}
            \caption{\label{fig: voronoi areas}Average Voronoi tile surface, normalized by particle surface area $\pi a^2$. \textit{Metastable} deposits are in red, and \textit{stabilized} deposits are in blue. }
        \end{figure}

        \medskip
        The preceding results indicate that interfacial cohesion is not governed solely by the number of inter-particle contacts, but also by the geometric constraints under which these contacts form upon deposition.
        This effect is most directly accessible in the first layer, where planar confinement by the substrate provides a well-defined reference for in-layer structural organization, which in turn constrains the orientation of contacts formed with the overlying layer.
        To probe how aggregation constrains local organization, we analyze the two-dimensional structure of the first layer via Voronoi tessellation (Figure~\ref{fig: voronoi diag}).
        To this end, particles belonging to the first layer are first selected using the $1$-$2$ inter-layer boundary, located at $z=1.57a$ and $z=1.31a$ for \textit{metastable} and \textit{stabilized} deposits, respectively. 
        Their positions are then projected onto the $xy$ plane, where each particle is assigned its Voronoi tile, defined as the set of all points closer to the particle's center than to that of any other particle.

        For \textit{stabilized} deposits, particles are consistently arranged in locally compact crystalline domains embedded in a more disordered background. Predominantly hexagonal, the crystalline regions also include square arrangements compatible with an overlying fcc packing, in agreement with previous observations made on similar deposits simulated by Giera \textit{et al.} \cite{giera_mesoscale_2017}.
        At high $E$, \textit{metastable} deposits exhibit first-layer organization similar to \textit{stabilized} deposits, consistent with the convergence of deposition regimes discussed above. At lower $E$, however, crystalline domains become markedly less prevalent in the first layer of \textit{metastable} deposits.  
        
        Figure~\ref{fig: voronoi areas} reports on the evolution with $E$ of the average Voronoi tile area, a way to quantify the trends observed on tilings.
        While the minimum conceivable area corresponds to that of a regular hexagon circumscribed around a particle ($\mathrm{A_{hex}}$ line on the graph), none of the systems investigated form a monocrystalline first layer, and the measured averages therefore lie well above this ideal value.
        For \textit{stabilized} deposits, the average Voronoi area shows no significant dependence on the electric field. In contrast, \textit{metastable} deposits display a pronounced increase in Voronoi area upon reducing the electric field, providing a clear quantitative signature of structural disorder within the first layer, and likely in the subsequent layers as well.
        This behavior strongly suggests that aggregation plays a central role in disrupting local packing during the early stages of deposition.
        Deposition proceeds through a sequence of successive layer-formation stages that partially overlap in time.
        Particle insertion into a given layer relies on the internal rearrangement dynamics under the mechanical constraints locally exerted by overlying particles.
        The static analysis of the interfacial region presented here indicates that aggregation significantly alters this insertion process, although the microscopic mechanisms by which it does so remain unresolved.
        This limitation highlights the need for an investigation of the deposition dynamics to fully capture the microscopic processes governing interfacial structure formation.

\section{\label{conclusion}Conclusion}
    In this work we tuned the particle simulation of electrophoretic deposition as modeled by Giera~\textit{et al}~\cite{giera_mesoscale_2017}, representing exclusively non-aggregating particles, by incorporating a classical DLVO potential, featuring a deep potential well, giving the particles the ability to irreversibly bond upon close approach.
    We isolated the effect of aggregation on the formation of the deposit and identified its microstructural signatures, in a drift-dominant regime.
    For that purpose, we compared the aggregating test system to a control system, identical to it except that the attractive interaction has been removed.
    These systems were subjected to a range of intensities of electric fields and a static analysis of the final microstructure was performed.

    We observed a clear influence of aggregation in the microstructure for electric fields below 6~MV~m$^{-1}$.
    Indeed, above this value the microstructure of deposits from \textit{metastable} and \textit{stabilized} particles seem to converge to the same behavior, which we interpreted as a transition in the deposition regime.
    In the core of the deposit, the aggregation tends to promote deposits of lower density and higher porosity.
    It disorganizes the orientation statistics of the contact network between layers, making it more isotropic, while increasing the inter-layer connectivity.
    For \textit{metastable} deposits, this translates into a tendency for enhanced inter-layer tensile response, although the statistical significance of this effect remained limited in our data.
    The maximum core packing fraction we observed, $\phi \approx 0.63$, significantly lower than that of a compact structure, as well as the negative density gradient we observed along the core of both \textit{metastable} and \textit{stabilized} systems, point toward the following threefold mechanism governing deposition in the core. 
    At the deposition front, rapid densification occurs simultaneously with local rearrangements, until a kinetic arrest freezes the configuration due to crowding effects. 
    The structure then experiences a slow relaxation under hydrostatic load, akin to glass aging.

    The structure of the deposit keeps a record of the configuration frozen at kinetic arrest, a record that slowly degrades over time as aging affects the structure.
    The aggregation likely influences the mechanisms of front densification, kinetic arrest as well as slow relaxation.

    The present work focuses on the mechanical signature within our model system; however, the same methodology could be applied to experimentally relevant electric fields.
    However, microstructure alone is insufficient to draw detailed mechanistic conclusions with high confidence.
    On the one hand, analysis of deposition dynamics in a frame co-moving with the front could complement this work and help attribute some structural signatures to specific stages of deposition.
    On the other hand, time analysis of layer properties could provide a natural framework to investigate the deposition mechanisms that occur specifically at the interface.
    More broadly, in the context of refining the model in order to get closer to experimentally relevant questions, it is crucial to accelerate the simulation protocol in order to be able to access much longer time scales, and thus studying the deposition under realistic electric fields.
    Finally, many parameters affect the deposition process, $\zeta$-potential or suspension dispersity for instance. 
    The method could be used to investigate their structural and mechanistic influence on the formation of EPD deposits.
    
\section*{Acknowledgment}
Rémi Martin thanks the French «Ministère de l’Enseignement Supérieur et de la Recherche et de l'Espace (MESRE)» for his PhD grant. This work was granted access to the HPC resources of the CALMIP supercomputing centre under the allocation P21014. The authors acknowledge Florence Ansart, Pierre-Louis Taberna, Vincent Dahirel and Marie Jardat for useful discussions.




\section*{Author Contributions}

Rémi Martin: Methodology, Software, Validation, Investigation, Formal analysis, Writing – Original Draft, Writing – Review \& Editing. Sandrine Duluard: Methodology, Writing – Review \& Editing, Supervision, Resources, Funding acquisition. Céline Merlet: Conceptualization, Formal analysis, Methodology, Validation, Writing – Review \& Editing, Supervision, Resources, Funding acquisition.

\section*{Conflicts of interest}

There are no conflicts to declare.
  
\nocite{*}

\providecommand{\noopsort}[1]{}\providecommand{\singleletter}[1]{#1}%

\end{document}